\documentclass[twocolumn,showpacs,preprintnumbers,superscriptaddress,amsmath,amssymb]{revtex4-2}
\usepackage{amssymb}		
\usepackage{amsmath}
\usepackage{graphicx}
\usepackage[normalem]{ulem}
\usepackage{multirow}
\usepackage{appendix}
\usepackage{CJK}
\usepackage[usenames]{color}
\usepackage{bm}
\usepackage[colorlinks,linkcolor=blue,urlcolor=blue,citecolor=blue]{hyperref}
\usepackage{booktabs} 	
\usepackage{leftidx}

\usepackage{soul,xcolor}
\setstcolor{red}

\usepackage[all]{hypcap} 
\makeatletter

\newcommand{\Rmnum}[1]{\expandafter\@slowromancap\romannumeral #1@}
\makeatother

\newcommand\redsout{\bgroup\markoverwith{\textcolor{red}{\rule[0.5ex]{2pt}{0.4pt}}}\ULon}

\usepackage{tikz,xcolor,hyperref}

\definecolor{lime}{HTML}{A6CE39}
\DeclareRobustCommand{\orcidicon}{
	\begin{tikzpicture}
	\draw[lime, fill=lime] (0,0) 
	circle [radius=0.16] 
	node[white] {{\fontfamily{qag}\selectfont \tiny ID}};
	\draw[white, fill=white] (-0.0625,0.095) 
	circle [radius=0.007];
	\end{tikzpicture}
	\hspace{-2mm}
}
\foreach \x in {A, ..., Z}{%
	\expandafter\xdef\csname orcid\x\endcsname{\noexpand\href{https://orcid.org/\csname orcidauthor\x\endcsname}{\noexpand\orcidicon}}
}

\begin{document}
\begin{CJK*}{UTF8}{gbsn}


\title{Impact of nuclear structure on the CME background in $^{96}_{44}$Ru + $^{96}_{44}$Ru and $^{96}_{40}$Zr + $^{96}_{40}$Zr collisions at $\sqrt{s_{NN}}$ = 7.7 $\sim$ 200 GeV from a multiphase transport  model}

\author{Fei Li(李飞)}
  \affiliation{Key Laboratory of Nuclear Physics and Ion-beam Application (MOE), Institute of Modern Physics, Fudan University, Shanghai 200433, China}
  
\author{Yu-Gang Ma(马余刚)\orcidB{}}%
 \email{mayugang@fudan.edu.cn}
  \affiliation{Key Laboratory of Nuclear Physics and Ion-beam Application (MOE), Institute of Modern Physics, Fudan University, Shanghai 200433, China}
  \affiliation{  Shanghai Research Center for Theoretical Nuclear Physics， NSFC and Fudan University, Shanghai 200438, China}

  \author{Song Zhang(张松)\orcidC{}}
  \affiliation{Key Laboratory of Nuclear Physics and Ion-beam Application (MOE), Institute of Modern Physics, Fudan University, Shanghai 200433, China}
    \affiliation{  Shanghai Research Center for Theoretical Nuclear Physics， NSFC and Fudan University, Shanghai 200438, China}

 \author{Guo-Liang Ma(马国亮)\orcidD{}}
  \affiliation{Key Laboratory of Nuclear Physics and Ion-beam Application (MOE), Institute of Modern Physics, Fudan University, Shanghai 200433, China}
    \affiliation{  Shanghai Research Center for Theoretical Nuclear Physics， NSFC and Fudan University, Shanghai 200438, China}

      \author{Qiye Shou(寿齐烨)\orcidE{}}
   \email{shouqiye@fudan.edu.cn}
  \affiliation{Key Laboratory of Nuclear Physics and Ion-beam Application (MOE), Institute of Modern Physics, Fudan University, Shanghai 200433, China}
    \affiliation{  Shanghai Research Center for Theoretical Nuclear Physics， NSFC and Fudan University, Shanghai 200438, China}
     
\date{\today}

\begin{abstract}
Impacts of nuclear structure on multiplicity ($N_{ch}$) and anisotropic flows ($v_{2}$ and $v_{3}$) in the isobaric collisions of $^{96}_{44}$Ru + $^{96}_{44}$Ru and $^{96}_{40}$Zr + $^{96}_{40}$Zr at $\sqrt{s_{NN}}$ = 7.7, 27, 62.4 and 200 GeV are investigated by using the string melting version of A MultiPhase Transport (AMPT) model. 
In comparison with  the experimental data released recently by the STAR collaboration, it is found that the impact of quadrupole deformation $\beta_{2}$ on the $v_{2}$ difference is mainly manifested in the most central collisions, while the octupole deformation $\beta_{3}$ is in the near-central collisions, 
and the neutron skin effect dominates in the mid-central collisions.  Viewing from the energy dependence, these effects are magnified at lower energies.
\end{abstract}
\maketitle


\section{Introduction}

 The present theories predict that  a local parity $(\textit{P})$ and charge parity $(\textit{CP})$ violation region could be formed by strong interaction in relativistic heavy-ion collisions~\cite{PhysRevLett.81.512, PhysRevD.61.111901, morley1985strong,abdallah2021search}, where  a charge number imbalance of light quarks can be achieved.  In the process of a non-central heavy-ion collision, the \textit{CP}-violating region is affected by the strong magnetic field produced by high speed protons passing through \cite{skokov2009estimate,PhysRevC.99.064607}, resulting in charge separation along the magnetic field. This phenomenon is also known as the Chiral Magnetic Effect (CME)~\cite{PhysRevD.78.074033,kharzeev2008effects,kharzeev2006parity,LiuYC,GaoJH}.
 The confirmation of the existence of the CME will not only lead to a deeper understanding for QCD vacuum, but also imply the existence of the strong-interaction \textit{CP}-violating region as well as the restoration of the chiral symmetry in Quark Gluon Plasma (QGP). This phenomenon can be attributed by the influence of strongest known electromagnetic field on the collision region~\cite{PhysRevD.61.111901,kharzeev2021chiral,kharzeev2016chiral} etc. Various anomalous chiral phenomena and possible detectable methods have been discussed in literature, e.g.,  \cite{kharzeev2016chiral,Fang_CPL,HuangXG_CPL,WangFQ_NST,Peng_CPL,LiW,Zhao_ppnp,Ren,Chang}. 
 Finding an experimental signature that can conclusively confirm the CME is one of current major challenges in relativistic heavy-ion physics.
 
 Since the magnetic field is usually perpendicular to the reaction plane ($RP$),  defined by the  impact parameter and the beam momentum in heavy-ion collisions, the CME-sensitive charge separation is measured with respect to the reaction plane, and the most widely used observable at present is the ``$\gamma$  correlator": $ \gamma_{\alpha \beta} = \left \langle cos(\phi_{\alpha} + \phi_{\beta} - 2 \Psi_{RP}  ) \right \rangle $ ~\cite{PhysRevC.70.057901}, where $\phi_{\alpha}$ and $\phi_{\beta}$  are the azimuthal angle of charged particles of interest, and $\Psi_{RP}$ is the angle of the reaction plane. However, some non-CME signal sources (e.g. local charge conservation and/or transverse momentum conservation entwined with the elliptic flow, $v_{2}$) can also contribute to the $\gamma$ value, which makes it difficult to quantify the CME effect in this way~\cite{PhysRevC.83.014905,PhysRevC.82.054902,PhysRevC.81.064902}.
 
 In order to disentangle the contribution of non-CME background to $\gamma$, many ideas have been proposed, among which the isobar collisions
 (e.g. $^{96}_{44}$Ru + $^{96}_{44}$Ru and $^{96}_{40}$Zr + $^{96}_{40}$Zr)
 are expected to provide the best solution to this problem. 
Since ${}^{96}_{44}$Ru +${}^{96}_{44}$Ru and ${}^{96}_{40}$Zr + ${}^{96}_{40}$Zr have the same nucleon numbers but  different charges, it is expected that different CME strengths can be quantitatively extracted from similar flow-driven backgrounds ~\cite{PhysRevC.94.041901,tribedy2020status}.  
 The experimental project of $^{96}_{44}$Ru + $^{96}_{44}$Ru and $^{96}_{40}$Zr + $^{96}_{40}$Zr isobar collisions at $\sqrt{s_{NN}}$ = 200 GeV was launched at RHIC in 2018, and the results were recently released  by  the STAR collaboration~\cite{abdallah2021search}.
 The isobar blind analysis 
 in the STAR collaboration \cite{NST} observed significant differences in the multiplicity
 and flow harmonics in a given centrality between the two collision systems, indicating that the magnitude of the
 non-CME background is inconsistent between the two species. 
 Many studies suggested that the discrepancy of CME backgrounds in Ru and Zr collisions was due to the different geometrical shapes  of colliding ions, such as the difference of deformation and neutron skin thickness for Ru and Zr~\cite{PhysRevC.94.041901,zhang2021evidence,PhysRevLett.121.022301}.
 
 In this paper, the CME background difference between Ru and Zr collisions is studied  in A MultiPhase Transport (AMPT) model, with five configurations for the nuclear structure parameters of Ru and Zr. Our results are compared with the experimental data to verify which description for nucleon structure of Ru and Zr is more consistent with the realistic nuclear structure.
 Further, the energy dependence of the CME background differences between Ru and Zr is investigated at various energies, i.e.  $\sqrt{s_{NN}}$ = 7.7, 27, 62.4 and 200 GeV, aiming at probing the nature of the background at lower energies and providing theoretical support for future experiments.
 
 The paper is organized as follows: In Sect. 2, the general setup of modeling Ru + Ru and Zr + Zr collisions by AMPT is  briefly introduced. In Sect. 3, the numerical results and discussion are presented, and the summary is given in Sec. 4.

\begin{figure*}
 	\centering
 	\includegraphics[scale=.9]{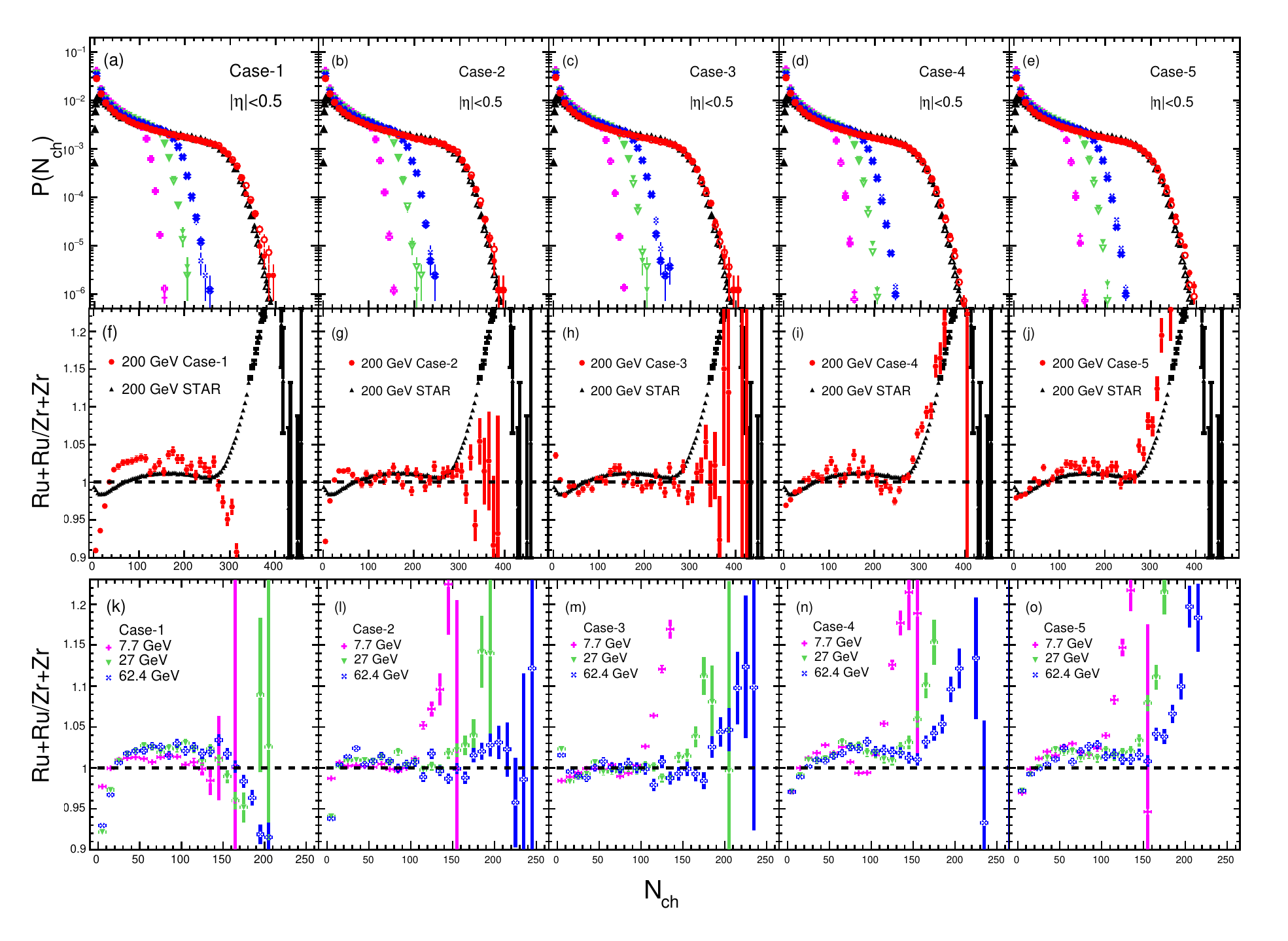}
 	\caption{(Upper panels) Distributions of the number of charged hadrons from the simulation of AMPT in the pseudorapidity window $|\eta|<$ 0.5 in Ru + Ru and Zr + Zr collisions at $\sqrt{s_{NN}}$ = 200 GeV for five sets of Woods-Saxon parameters: (a) Case-1, (b) Case-2, (c) Case-3, (d) Case-4, and (e) Case-5. More details on definition of cases can be found in texts.  The STAR data~\cite{abdallah2021search} for isobar collisions at $\sqrt{s_{NN}}$ = 200 GeV  are also shown for comparisons. (Middle panels) The Ru + Ru to Zr + Zr ratio for five sets of Woods-Saxon parameters at $\sqrt{s_{NN}}$ = 200 GeV, as well as the STAR data are shown, respectively.
 	(Lower panels) The Ru + Ru to Zr + Zr ratio for five sets of Woods-Saxon parameters at $\sqrt{s_{NN}}$ = 7.7, 27 and 62.4 GeV, respectively. Notice that the horizontal axis is zoomed in for clarity in the lower pannels.
 	}
 	\label{FIG:mul_ntrack}
 \end{figure*}
 
\section{General Setup}

\subsection{The AMPT model}
 
Here we employed the string melting version of the AMPT model~\cite{PhysRevC.72.064901,AMPT2021} to simulate Ru + Ru and Zr + Zr collisions and analyzed the generated data to study the CME backgrounds. The model  has proven to be effective in describing collective flow in small and large collision systems at RHIC and LHC ~\cite{PhysRevC.94.054910, MA2014209, PhysRevLett.113.252301,  PhysRevC.98.034903,WangH}.
Based on the nonequilibrium transport dynamics, the AMPT model is composed of four parts: the Heavy-Ion Jet INteraction Generator (HIJING) model~\cite{PhysRevD.44.3501, GYULASSY1994307} for generating the initial-state information, 
Zhang's parton cascade (ZPC) model~\cite{ZHANG1998193} for modeling partonic scatterings, the Lund string fragmentation model or a quark coalescence model for hadrons formation, and a relativistic transport (ART) model~\cite{PhysRevC.52.2037} for treating the hadron scatterings. 
In the AMPT model with string melting mechanism, the partons freeze-out is according to local energy density and the hadronization process is simulated by a naive quark coalescence model, which combines two (three) nearest partons into a meson (baryon). 
The method for determining hadronic species is achieved by the flavor and invariant mass of coalescing partons.

\begin{figure*}
	\centering
	\includegraphics[scale=.9]{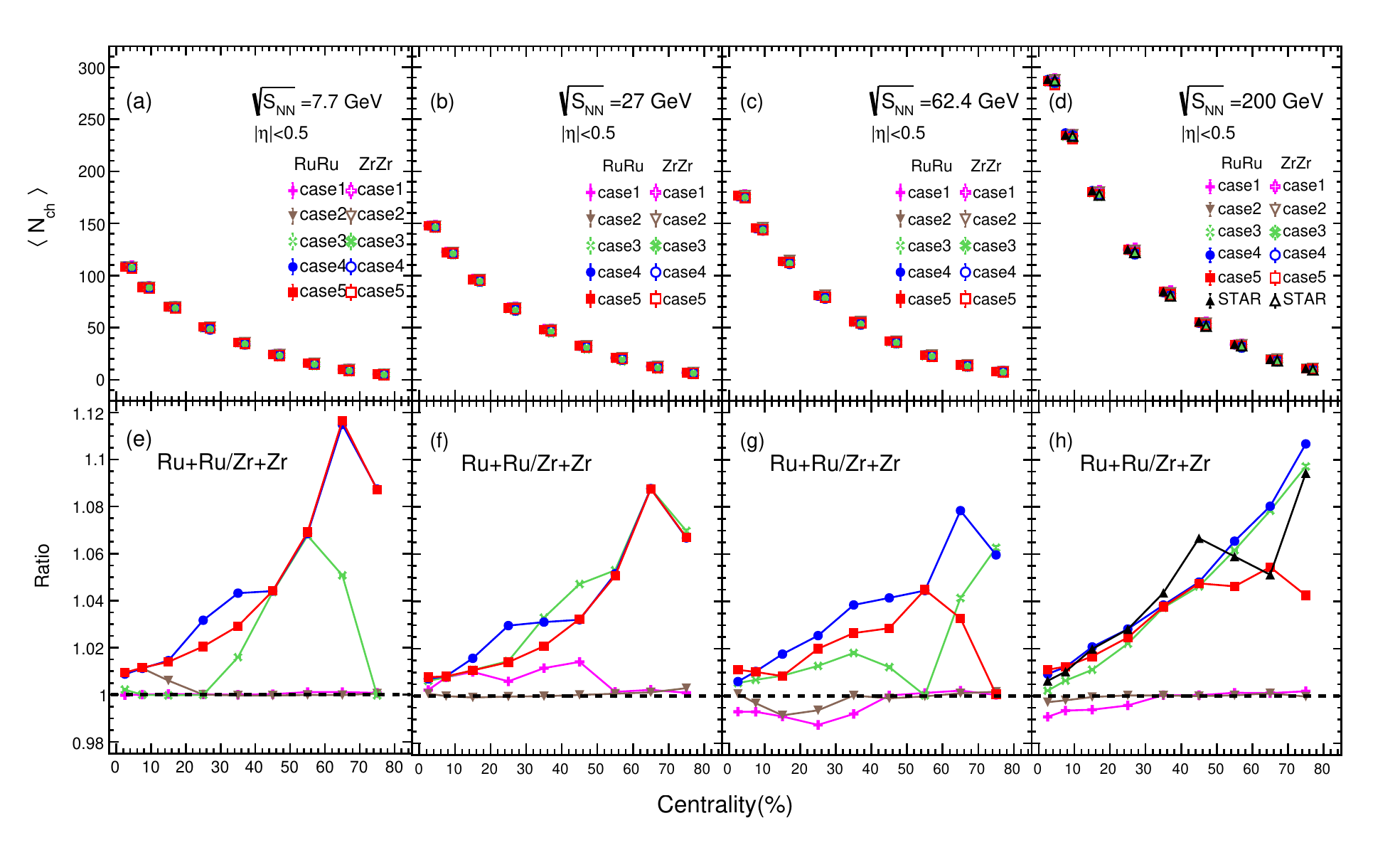}
	\caption{(Upper panels) The mean charge multiplicity $N_{ch}$ within $|\eta| <$ 0.5 as a function of centrality in Ru + Ru and Zr + Zr collisions at $\sqrt{s_{NN}}$ = 7.7, 27, 62.4 and 200 GeV.  The STAR data~\cite{abdallah2021search} for isobar collisions at $\sqrt{s_{NN}}$ = 200 GeV  are also shown for comparison. The centrality bins are slighted shifted for clarity.  (Lower panels) The ratio of the mean charge multiplicity in Ru + Ru collisions to that in Zr + Zr collisions in matching centrality and energy. The above data include statistical uncertainty. }
	\label{FIG:mul}
\end{figure*}

\subsection{Description of $^{96}_{44}$Ru and $^{96}_{40}$Zr}
 
The spatial distribution of nucleons in the rest frame of $^{96}_{44}$Ru and $^{96}_{40}$Zr  can be described by the following 2-parameter Fermi mass density of the Woods-Saxon (WS) form~\cite{PhysRevC.94.041901,PRL2021,ShouPLB}: 
\begin{equation}
 \rho ( r , \theta ) = \frac { \rho _ { 0 } } { 1 + \operatorname { exp }\left [ \frac{  r - R _ { 0 } \left(1+ \beta _ { 2 } Y _ { 2 } ^ { 0 } ( \theta ) + \beta _ { 3 } Y _ { 3 } ^ { 0 } ( \theta  \right) } {a} \right] },
 \end{equation}
 where \textit{r} is radial position and $\theta$ is polar angle in spherical coordinates, $\rho_{0}$ = 0.16 fm$^{-3}$ is the nuclear saturation density, $R_{0}$ and $a$  represent the “radius” of the nucleus and the surface diffuseness parameter, respectively, and the deformation of the nucleus is denoted by the most relevant axial symmetric quadrupole deformation $\beta_2$ and octupole deformation $\beta_3$.  Since the  $\beta_{2}$ and $\beta_{3}$ values of $^{96}_{44}$Ru and $^{96}_{40}$Zr are not accurately known at  present~\cite{PhysRevC.94.041901, PhysRevC.98.054907,PhysRevC.101.061901},
 here we take five sets of WS parameters which are suggested and used in heavy-ion collisions from recent references~\cite{RAMAN20011,PRITYCHENKO20161,MOLLER1995185,XU2021136453, jia2021scaling, nijs2021inferring}, as shown in Table~\ref{tbl1}, to investigate the effect of nuclear structures on CME backgrounds for isobar collisions.
 
 The parameters are arranged by the following way: Case-1, Ru ($\beta^{Ru}_{2}$ = 0.13)  has large quadrupole deformation than Zr ($\beta^{Zr}_{2}$ = 0.06); Case-2, Ru ($\beta^{Ru}_{2}$ = 0.03), in contrast, has a smaller quadrupole deformation than Zr($\beta^{Zr}_{2}$ = 0.18); 
 Case-3 is based on the latest calculations of the energy density functional theory (DFT)~\cite{XU2021136453,PhysRevLett.121.022301}, which assumes that the nucleus is spherical ($\beta_{2}$ = 0). According to the calculation of proton and neutron distribution, it shows that the overall size of Ru is smaller than that of Zr because the neutron skin of Zr is much thicker. 
Case-4 is also derived from the DFT result, but with recent research findings~\cite{jia2021scaling,nijs2021inferring} for $\beta_2$ and $\beta_3$, which gives a description of nuclear structure  contained both the deformation effect and the neutron skin effect for $^{96}_{44}$Ru and $^{96}_{40}$Zr.
Case-5 is from the recent result~\cite{jia2021scaling}, which also includes the deformation effect and the neutron skin effect.
It should be noted that the neutron skin effect comes from the different density distribution of proton and neutron, but in our computation, proton and neutron are identical in Case-3, Case-4 and Case-5, and the total nucleon density (the sum of the proton and neutron) is used to configure the initial nucleon coordinates.
It has been proved in previous research~\cite{XU2021136453} that the Ru+Ru/Zr+Zr ratio is almost the same for charged multiplicity distribution $N_{ch}$ and the eccentricity $\varepsilon_{2}$ which were calculated by the total nucleon density and the DFT density. This makes it possible to study the neutron skin effect by using the WS parameters of the total nucleon density. It has been proved in a recent work~\cite{jia2021scaling} that the effects of WS parameters on elliptic flow ratio of $v_{2,Ru}$/$v_{2,Zr}$ are as follows: 1) the $v_{2}$ ratio is mainly dominated by $\beta_{2}$ and to a minor extent by $\beta_{3}$ in the most-central collisions; 2) the $v_{2}$ ratio is influenced by a positive contribution from $\beta_{2}$ and a larger negative contribution from $\beta_{3}$ in the near-central collisions; 3) the influence of $\Delta a = a_{Ru}-a_{Zr}$ is manifested in the mid-central and peripheral collisions; 4) and the impact of $\Delta R_{0} = R_{0,Ru}-R_{0,Zr}$ only exists in the central collisions.

  \begin{table}[hbp]
 	\centering
 	\caption{The Woods-Saxon parameters used in the AMPT model.}
 	\label{tbl1}
 	\fontsize{9}{20}
 	\renewcommand\arraystretch{1.2}
 	\begin{tabular}  {|c|c|c|c|c|c|c|c|c|}
 		\hline
 		 &\multicolumn{4}{c|}{$^{96}_{44}$Ru}  &\multicolumn{4}{c|}{$^{96}_{40}$Zr}\cr\cline{2-9}
 Nucleus		 		& $R$(fm) &  $a$(fm)   &  $\beta_{2}$   &  $\beta_{3}$  
 		& $R$(fm) &  $a$(fm)  &  $\beta_{2}$   &  $\beta_{3}$   \cr
 		\hline
Case-1
 		& 5.13&  0.46& 0.13& 0 
 		& 5.06&  0.46 & 0.06& 0 \cr\cline{2-9}\hline
Case-2
 		& 5.13&  0.46 & 0.03& 0 
 		& 5.06&  0.46 & 0.18& 0 \cr\cline{2-9}\hline
Case-3
 		& 5.067&  0.5&  0& 0
 		& 4.965 & 0.556 & 0& 0 \cr\cline{2-9}\hline
Case-4
 		& 5.065&  0.485&  0.154&  0
 		& 4.961 & 0.544 & 0.062&  0.202 \cr\cline{2-9}\hline
Case-5
 		& 5.09&  0.46&  0.162&  0
 		& 5.02 & 0.52 & 0.06&  0.20 \cr\cline{2-9}\hline
 	\end{tabular}
 \end{table}
 
Figure~\ref{FIG:mul_ntrack} shows the distributions of charged hadron numbers from the AMPT model  within $|\eta| <$ 0.5 in Ru + Ru and Zr + Zr collisions at $\sqrt{s_{NN}}$ = 7.7, 27, 62.4 and 200 GeV for above five sets of Woods-Saxon parameters. 
 The multiplicity $P(N_{ch})$ of five cases perfectly matches the STAR multiplicity after multiplying by 1.21 (the same treatment was done  in the work ~\cite{nijs2021inferring}) at $\sqrt{s_{NN}}$ = 200 GeV, and the ratio of the Case-4 and the Case-5 give the best description of the RHIC-STAR data~\cite{abdallah2021search}.  It presents similar distribution pattern at other energies in different $N_{ch}$ region. Based on the multiplicity distribution, the collision system is divided into different centralities for each set of the WS parameters. It's obvious that the $N_{ch}$ interval of centralities class becomes more sensitive to the distribution for low multiplicity (peripheral collisions) at lower energy.  Here the multiplicity $N_{ch}$ is the number of particles at mid-pseudorapidity $\left(  |\eta| < 0.5 \right)$ in the collisions.

\begin{figure*}
	\centering
	\includegraphics[scale=.85]{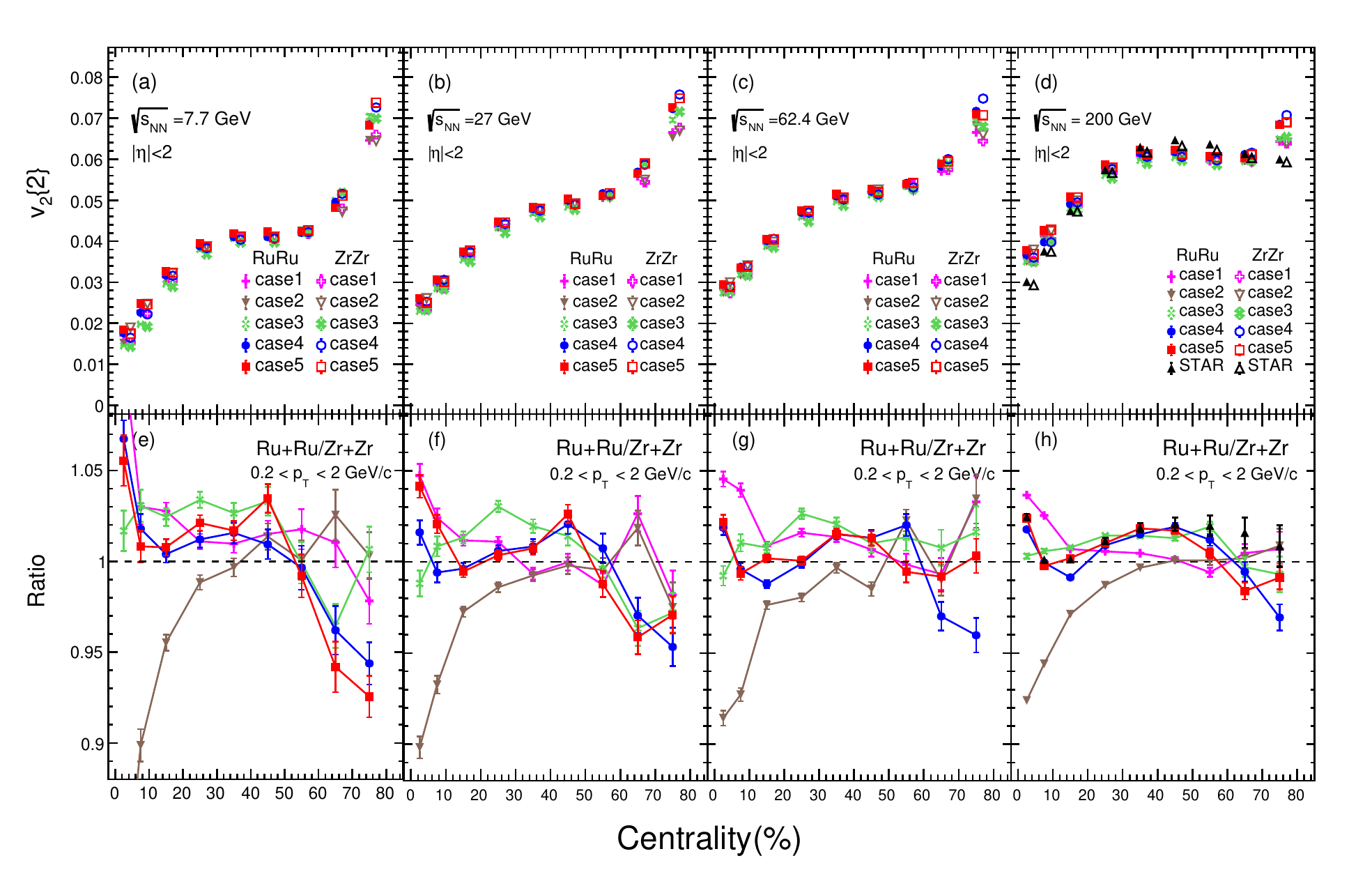}
	\caption{ (Upper panels) Elliptic flow $v_2\{2\}$  measurements for five cases in isobar collisions at $\sqrt{s_{NN}}$ = 7.7, 27, 62.4 and 200 GeV as a function of centrality by using two-particle correlations. The solid and open symbols represent measurements for Ru + Ru and Zr + Zr collisions, respectively.  The STAR data~\cite{abdallah2021search}  for isobar collisions at $\sqrt{s_{NN}}$ = 200 GeV  are also shown for comparison.
		The data points are shifted along the $x$ axis for clarity. (Lower panels) The ratios of $v_2$ in Ru + Ru over Zr + Zr collisions. The statistical uncertainties are represented by lines. }
	\label{FIG:v2all}
\end{figure*}

\section{RESULTS AND DISCUSSION}

In this section, the simulation results from the AMPT model for five nuclear density parameters of Ru + Ru and Zr + Zr collisions at $\sqrt{ s_{NN}}$ = 7.7, 27, 62.4 and 200 GeV are presented. We will show the predictions for the mean multiplicity $\left\langle N_{ch} \right\rangle$, the elliptic flow $v_{2}$ and the triangular flow  $v_{3}$  of all particles as a function of centrality in the two isobaric collision systems.
The effect of initial geometry on CME background will also be discussed according to the eccentricities in Ru + Ru and Zr + Zr collisions. 
The numbers of events we simulated for all cases for both $^{96}_{44}$Ru and $^{96}_{40}$Zr are, respectively,  5M  (7.7 GeV),  2M (27 GeV), 2M (62.4 GeV)  and 2.5M (200 GeV).

\subsection{Mean charge multiplicity $\left\langle N_{ch} \right\rangle$}

The upper panels of Fig. \ref{FIG:mul} show the mean charge multiplicity $\left\langle N_{ch} \right\rangle$ at mid-pseudorapidity $\left(  |\eta| < 0.5 \right)$ as a function of centrality, from the string melting mode of
AMPT, for five cases of the parameter settings of isobar collision systems at $\sqrt{ s_{NN}}$ = 7.7, 27, 62.4 and 200 GeV.
The Ru+Ru/Zr+Zr ratio of the mean charge multiplicities is shown in the lower panels of Fig. \ref{FIG:mul}.
The STAR data~\cite{abdallah2021search} for isobar collision systems at $\sqrt{s_{NN}}$ = 200 GeV are also shown for comparison.
Our simulation results show that the mean charge multiplicity $\left\langle N_{ch} \right\rangle$ of the two isobar systems for five cases is close to each other, which is perfectly consistent with the data of the STAR Collaboration at different centralities. 

The Ru+Ru/Zr+Zr ratio in the bottom panel of Fig. \ref{FIG:mul} gives a clear illustration of the difference between these two isobar collisions. The ratios for the five cases are different in shape, with Case-1 and Case-2 remaining close the unity for all centralities at different energies, Case-3, Case-4 and Case-5 rising almost directly in proportion to the centrality at $\sqrt{s_{NN}}$ = 200 GeV, which are also rising in a zig-zag pattern at low energy. The shapes of the ratio in the Case-3, Case-4 and Case-5 are the closest to the STAR measurements at $\sqrt{s_{NN}}$ = 200 GeV.
This indicates that the neutron skin effect has the major contribution to the ratio of $\left\langle N_{ch} \right\rangle$. 

One of the puzzling things about the $\left\langle N_{ch} \right\rangle$ ratio of STAR's data at $\sqrt{s_{NN}}$ = 200 GeV is that there is a zig-zag pattern after 50\% of the centrality, which is interestingly similar to how Case-3, Case-4 and Case-5 behaves at low energy.   The zig-zag ratio after 50\% of the centrality is actually due to the fact that in low multiplicity events (i.e., in peripheral collisions) the $N_{ch}$ interval for a specific centrality has a little difference between $^{96}_{44}$Ru and $^{96}_{40}$Zr \cite{nijs2021inferring}. In other word, the events in $^{96}_{44}$Ru and $^{96}_{40}$Zr collisions within the same centrality region have different multiplicity region. This little multiplicity difference can result in the zig-zag fluctuations for some observables.  As shown in Fig.~\ref{FIG:mul_ntrack}, the ratio with zig-zag fluctuations appears before 50\% of the centrality at low energy collisions, because low energy collisions contain more low multiplicity events.  In the work of Ref.~\cite{nijs2021inferring}, it points out that the ratio of multiplicity $\left\langle N_{ch} \right\rangle$ depends on centrality significantly, the ratio of $v_{2}$ dose not.

\subsection{Harmonic flow }

The harmonic flow is investigated for the two isobaric collision systems, which contributes as a major background of the CME-sensitive observable, such as $\Delta \gamma_{112}$.
The harmonic flow coefficient $v_n\{2\}$ is calculated by the two-particle correlation method in the following form~\cite{abdallah2021search}:
\begin{equation}
	\centering
v^{2}_{n=2,3}\{2\} = \left\langle cos(n\phi_{1} -  n\phi_{2})  \right\rangle .
\end{equation}
Here we use all particles and larger pseudo-rapidity range ($| \eta| < 2$) to calculate $v^{2}_n\{2\}$ for higher statistics,
and the $\Delta \eta_{1,2}$ is set to 0.05 for $v^{2}_n\{2\}$ calculation. The non-flow effect which can be suppressed by using a larger $\Delta \eta_{1,2}$ cut does not affect the ratio essentially~\cite{nijs2021inferring}.

The upper panels of Fig.~\ref{FIG:v2all} present the AMPT results for the centrality dependence of elliptic flow $v_{2}$\{2\} at mid-pseudorapidity ($| \eta | < $2) with five geometry settings of isobaric collisions.
It can be seen that $v_{2}$\{2\}  for Ru + Ru and Zr + Zr are similar to each other at different energies, and the  $v_{2}$\{2\} with five setting parameters are also close to each other. Our results are consistent with the STAR data at $\sqrt{ s_{NN}}$ = 200~GeV, but slightly higher than that of the STAR data in central collisions.

\begin{figure}
	\centering
	\includegraphics[scale=.45]{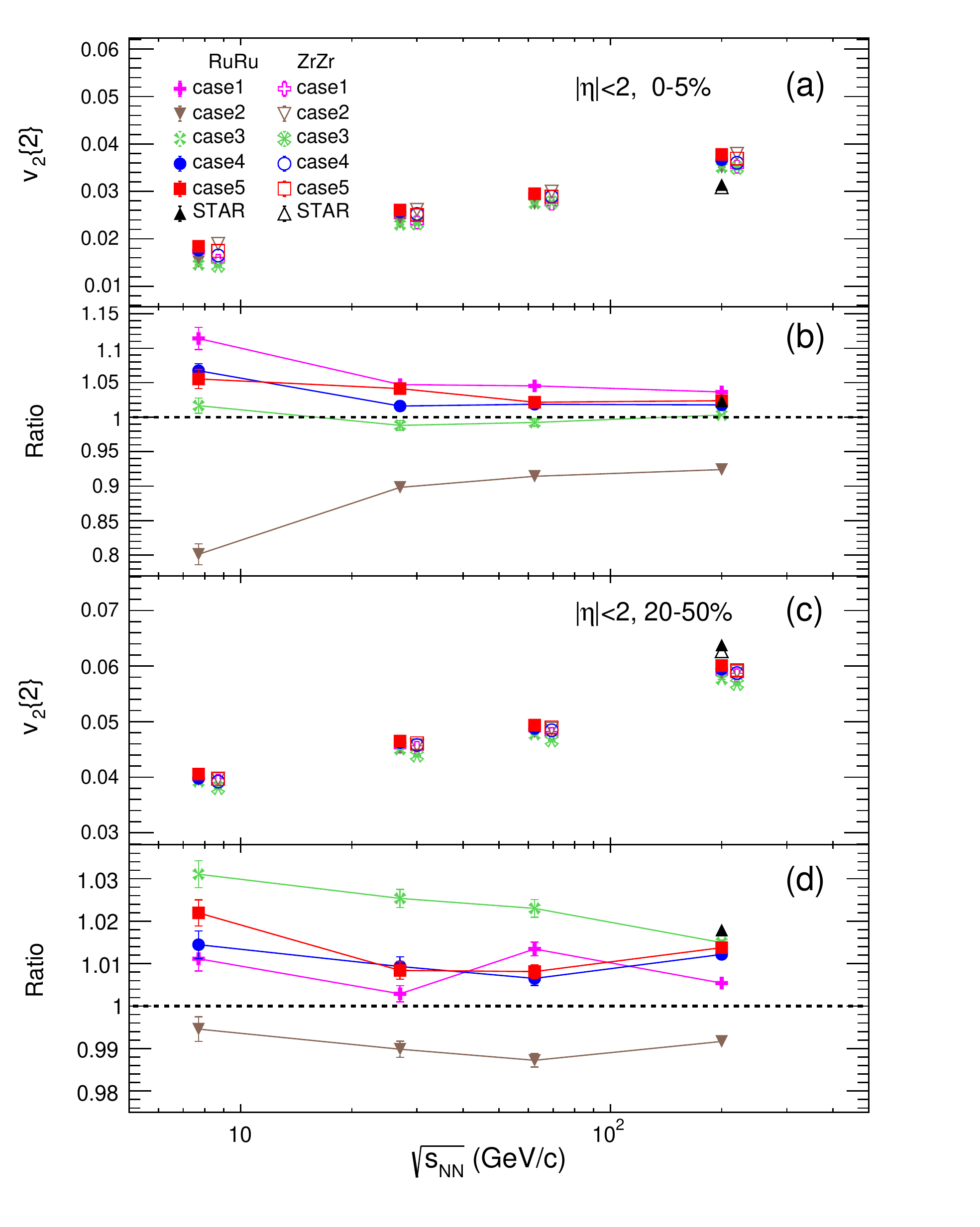}
	\caption{ (a) Elliptic flow $v_2\{2\}$ measurements for five cases in the most central isobar collisions as a function of energy by using two-particle correlations. (b) The ratios of $v_2\{2\}$ in Ru + Ru over Zr + Zr collisions in the most central collisions.  (c) Elliptic flow $v_2\{2\}$ measurements for five cases in mid-central isobar collisions as a function of energy by using two-particle correlations. (d) The ratios of $v_2\{2\}$ in Ru + Ru over Zr + Zr collisions in mid-central collisions. For (a) and (b), the STAR data~\cite{abdallah2021search}  for isobar collisions at $\sqrt{s_{NN}}$ = 200 GeV  are also shown for  comparison. The data points are shifted along the $x$ axis for clarity, and the statistical uncertainties are represented by lines.}
	\label{FIG:v12}
\end{figure}

The lower panels of Fig. \ref{FIG:v2all} give the $v_{2}$\{2\} ratios between Ru + Ru and Zr + Zr collisions. 
The centrality dependences of the ratios for the five settings are very different at $\sqrt{s_{NN}}$ = 200 GeV, for Case-1 (-2)  the ratio decreases (increases) from central to peripheral collisions until about 20\% centrality staying close to unit, and for Case-3 it shows as a bow above unit, and for Case-4 and Case-5 it shows the almost same shape as the STAR data.
As seen from Case-1 and Case-2, the quadrupole deformation has a major effect on the $v_{2}$\{2\} ratios in central (less than 20\% centrality) isobaric collision systems, which means that the larger quadrupole deformation $\beta_{2}$  of nucleon distribution  is, the larger its $v_{2}$\{2\}.
Comparing with  the  STAR experimental results~\cite{abdallah2021search}, Case-1 is more consistent with the data in the most central collisions, Case-3 is more consistent with it in mid-central collisions,
while Case-4 and Case-5, which take account of both the deformation and the neutron skin effect, are almost consistent with the STAR data in all centralities.  It can be seen that the influences of setting of the parameters are consistent with the conclusion in Ref.~\cite{jia2021scaling} and this also provokes the investigation of the beam energy dependence of the effect from the initial geometry properties.

\begin{figure}
	\centering
	\includegraphics[scale=.45]{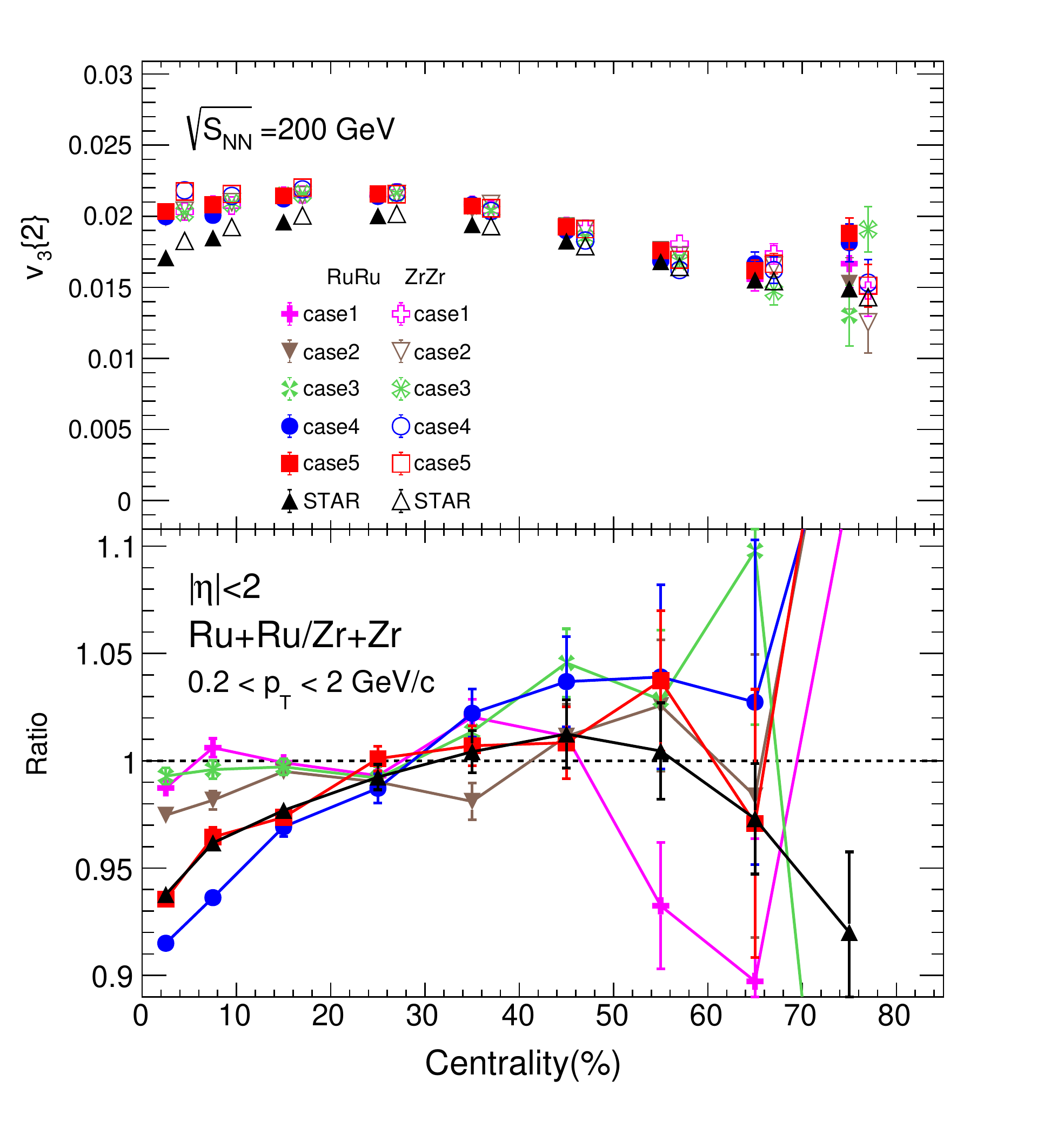}
	\caption{(Upper panels) Triangular flow $v_{3}$\{2\} for five cases in isobar collisions at $\sqrt{s_{NN}}$ = 200 GeV as a function of centrality by using two-particle correlations.  The STAR data~\cite{abdallah2021search}  for isobar collisions at $\sqrt{s_{NN}}$ = 200 GeV  are also shown for  comparison. The data points are shifted along the $x$ axis for clarity. (Lower panels) The ratios of $v_2$ in Ru + Ru over Zr + Zr collisions. The statistical uncertainties are represented by the bars on dots.}
	\label{FIG:v3}
\end{figure}

\begin{figure*}
	\centering
	\includegraphics[scale=.85]{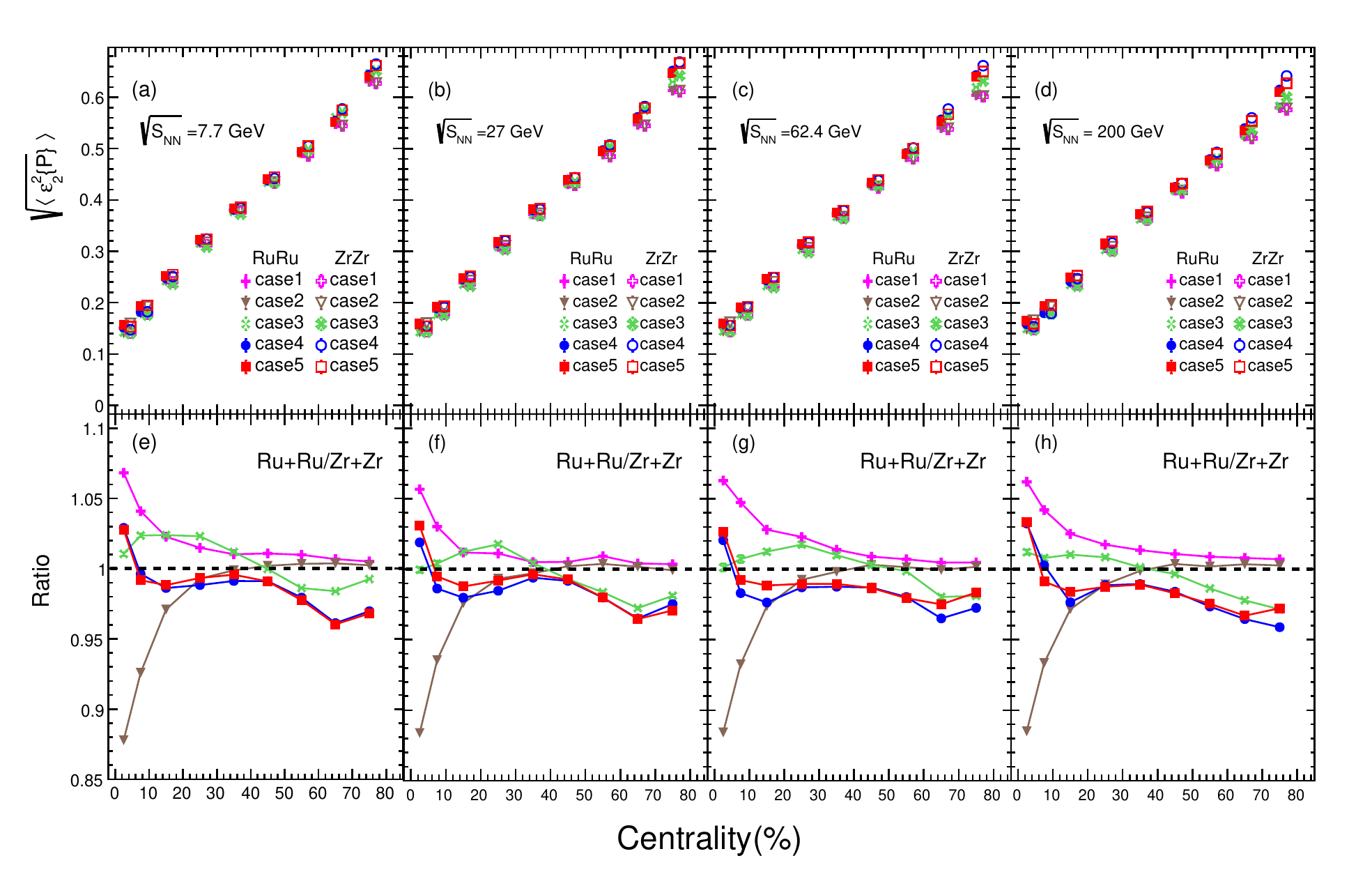}
	\caption{ (Upper panels) Eccentricity $\varepsilon_{2}$ for three cases in isobar collisions at $\sqrt{s_{NN}}$ = 7.7, 27, 62.4 and 200 GeV as a function of centrality. The solid and open symbols represent measurements for Ru + Ru and Zr + Zr collisions, respectively. The data points are shifted along the $x$ axis for clarity. (Lower panels) The ratios of $\varepsilon_{2}$ in Ru + Ru over Zr + Zr collisions. The statistical uncertainties are represented by bars on dots.}
		\label{FIG:ecc_all}
	\end{figure*}

It was also observed that the ratios of $v_{2}$\{2\} is enhanced at lower energies in isobar collisions, but the five settings have different performances at different centralities.
As shown in Fig. \ref{FIG:v12} (b), in the central collisions (0-5\%), the absolute value of $v_{2}$\{2\} ratios decreases with energy for Case-1, Case-2, Case-4 and Case-5，meaning that the effect of $\beta_{2}$ to $v_{2}$\{2\} ratio, which mainly dominates in the most-central collisions, is magnified at lower energy.
Comparing the ratios of these cases, with a larger absolute value of $\Delta \beta_{2,case2}$ ($\beta_{2,Ru}$ - $\beta_{2,Zr}$ = -0.15), the ratios of Case-2 drop more strongly in central region than other case, which indicates that a larger $\Delta \beta_{2}$ results in a larger $v_{2}$\{2\} ratio in central collisions. 
In addition, due to the negative contribution to $v_{2}$\{2\} ratio from $\beta_{3}$ in the near-central collisions, although the $\Delta \beta_2$ of Case-4 ($\Delta \beta_{2,case4}$ = 0.092) and Case-5 ($\Delta \beta_{2,case5}$ = 0.102) is larger than that of Case-1($\Delta \beta_{2,case1}$ = 0.07), their $v_{2}$\{2\} ratio is still smaller than that of Case-1.
It also can be seen that the ratio of Case-3, which does not include the deformation description in WS parameters, remains at unit with energy.

Figure \ref{FIG:v12}(d) shows the ratios of $v_{2}$\{2\} with energy in the mid-central collisions (20-50\%).
It can be seen that the ratio of Case-3 decreases with energy, while the other cases show a non-monotonic energy dependence of ratios. 
Since the neutron skin effect (i.e. $\Delta a$) is mainly manifested in the mid-central collisions, the behaviour of Case-3 in Fig.~\ref{FIG:v12}(d) indicate that the influence of neutron skin effect to the ratio of $v_{2}$\{2\} is enhanced at low energy in the mid-central collisions.
For Case-4 and Case-5, the ratios of $v_{2}$\{2\} shows a non-monotonic energy dependence in this collision energy region because both sets of WS parameters contain not only the neutron skin effect but also  octupole deformation $\beta_{3}$. Since in the near-central collisions $\beta_{3}$ has a negative contribution to $v_{2,Ru}$/$v_{2,Zr}$ and in the mid-central collisions $\Delta a$ has a positive contribution, the two factors conspire to produce the non-monotonic energy dependence of ratios in this centrality region (20-50\%).
The ratios of $v_{2}$\{2\} present a minor difference for Case-1 and Case-2 at different energy in this centrality region, because the quadrupole deformation $\beta_{2}$ has little effect on $v_{2}$\{2\} ratios in the mid-central collisions.

The trend of $v_{2}$\{2\} ratios at different energies with different centrality reconfirms that the effect of quadrupole deformation $\beta_{2}$ on $v_{2}$\{2\} ratios mainly occurs in the most-central collisions, the octupole deformation $\beta_{3}$ dominates in the near-central collisions, and the impact of neutron skin effect on $v_{2}$\{2\} ratios mainly happens in the mid-central collisions.

The above results tell us that each nuclear structural factor could result in a larger difference of CME background at low energy.
However, due to the negative contribution from $\beta_{3}$ to $v_{2,Ru}$/$v_{2,Zr}$  ratio in the near-central collisions, the monotonic energy dependence of $v_{2}$ ratio is weakened at 0-5\% and 20-50\%. 
Therefore, we suggest that the energy dependence of the ratio for isobar collisions can be studied in the extremely ultra-central collisions (0-1\%) \cite{nijs2021inferring} in the future, which can avoid the offset of the $\beta_{3}$ to $v_{2}$ ratio to obtain a stronger monotonic energy dependence.
At the same time, it can solve the problem that the centrality classes of two colliding nuclei are not completely aligned at low energy.

On the other hand, perhaps the lower energy collisions are more suitable for us to study the nuclear structure of ${}^{96}_{44}$Ru and ${}^{96}_{44}$Zr because of  stronger signal than that at higher energy. In other words, it is worth performing a beam energy scan of isobaric collisions to constrain the nuclear structure parameters or to disclose the CME background from nuclear structure.

We also investigate the effect of the quadrupole deformation $\beta_{2}$ and octupole deformation $\beta_{3}$ to the triangular flow $v_{3}$\{2\} at mid-pseudorapidity ($|\eta| <$ 2) as a function of centrality at $\sqrt{s_{NN}}$ = 200 GeV, as shown in Fig.~\ref{FIG:v3}.  
From the upper panels of Fig. \ref{FIG:v3}, it is seen that the calculated $v_{3}$\{2\} are close to the STAR data~\cite{abdallah2021search} for five settings, and the lower panels of Fig. \ref{FIG:v3} showing the results of Case-4 and Case-5, which include the octupole deformation $\beta_{3}$, are most similar to STAR data in central collisions.
The above results are consistent with the conclusion in Ref.~\cite{zhang2021evidence} that the $\beta_{2}$ has an effect on $v_2$ but does not on $v_3$, and $\beta_{3}$ has a significant effect on $v_{3}$\{2\} in central collisions. On the other hand, there exists a similar effect of the chain or triangle structure for the $\alpha$-clustering nucleus like $^{12}$C on $v_2$ and $v_3$,  which was observed in our previous work \cite{ZhangS}, i.e. the chain structure has  an effect on $v_2$ but does not on $v_3$, but triangle structure has a significant effect on $v_{3}$. Actually, both phenomena induced by the above  deformations and clustering structures are essentially related to the transformation from initial-state geometric structure and fluctuation \cite{MaL} to final-state momentum space.

\subsection{The initial geometry}

Next, we want to find  whether the differences in final state measurements due to the effects of nuclear structure are already implicit in the early stages after isobar collisions.
The initial geometry of a nucleus-nucleus collision can be characterized by eccentricity, which represents the initial geometric anisotropy of the collision zone in the transverse plane (perpendicular to the beam direction).
The definition of eccentricity for $n$-th harmonics in the coordinate space of the initial partons~\cite{EPJA.54.161-SZhang2018} for a single collision event is given by following form \cite{alver2008importance,adams2005azimuthal}
\begin{equation}
\varepsilon_{n}\{P\} = \frac{ \sqrt {  \langle r ^ { n }  cos ( n \varphi ) \rangle ^{2}+  \langle r ^ { n } \operatorname { sin } ( n \varphi ) \rangle^{2}   } }{\langle  r^{n}  \rangle },
\end{equation}
where $r$ and $\varphi$ are the position and azimuthal angle of each initial parton in the transverse plane. In practice, the event averaged eccentricity coefficients $\langle\varepsilon_{n}\{P\}\rangle$ are used to characterize the initial geometry asymmetry, and we mainly focus on the $\varepsilon_{2}$ in this study.

The upper panels of  Fig. \ref{FIG:ecc_all} show the $ \sqrt{\langle\varepsilon^{2}_{2}\rangle}$ for five cases in isobar collisions at $\sqrt{s_{NN}}$ = 7.7, 27, 62.4 and 200 GeV as a function of centrality, and the $\sqrt{\langle\varepsilon^{2}_{2}\rangle}$ for five settings are close to each other at different centralities. 
The lower panels of  Fig. \ref{FIG:ecc_all} present the $\sqrt{\langle\varepsilon^{2}_{2}\rangle}$ ratios between Ru + Ru and Zr + Zr collisions. As expected, the shape of $\sqrt{\langle\varepsilon^{2}_{2}\rangle}$ ratios is similar to  the above presented flow ratios for the five settings at different centralities, i.e. the effect of deformation on eccentricity mainly occurs in the central collisions, while the effect of neutron skin on eccentricity mainly dominates in mid-central collisions. 
This illustrates that part of the contribution to the difference on CME backgrounds in Ru + Ru and Zr + Zr collisions originates from the effect of nuclear structure on the initial geometry after collisions.

	\begin{figure}
	\centering
	\includegraphics[scale=.4]{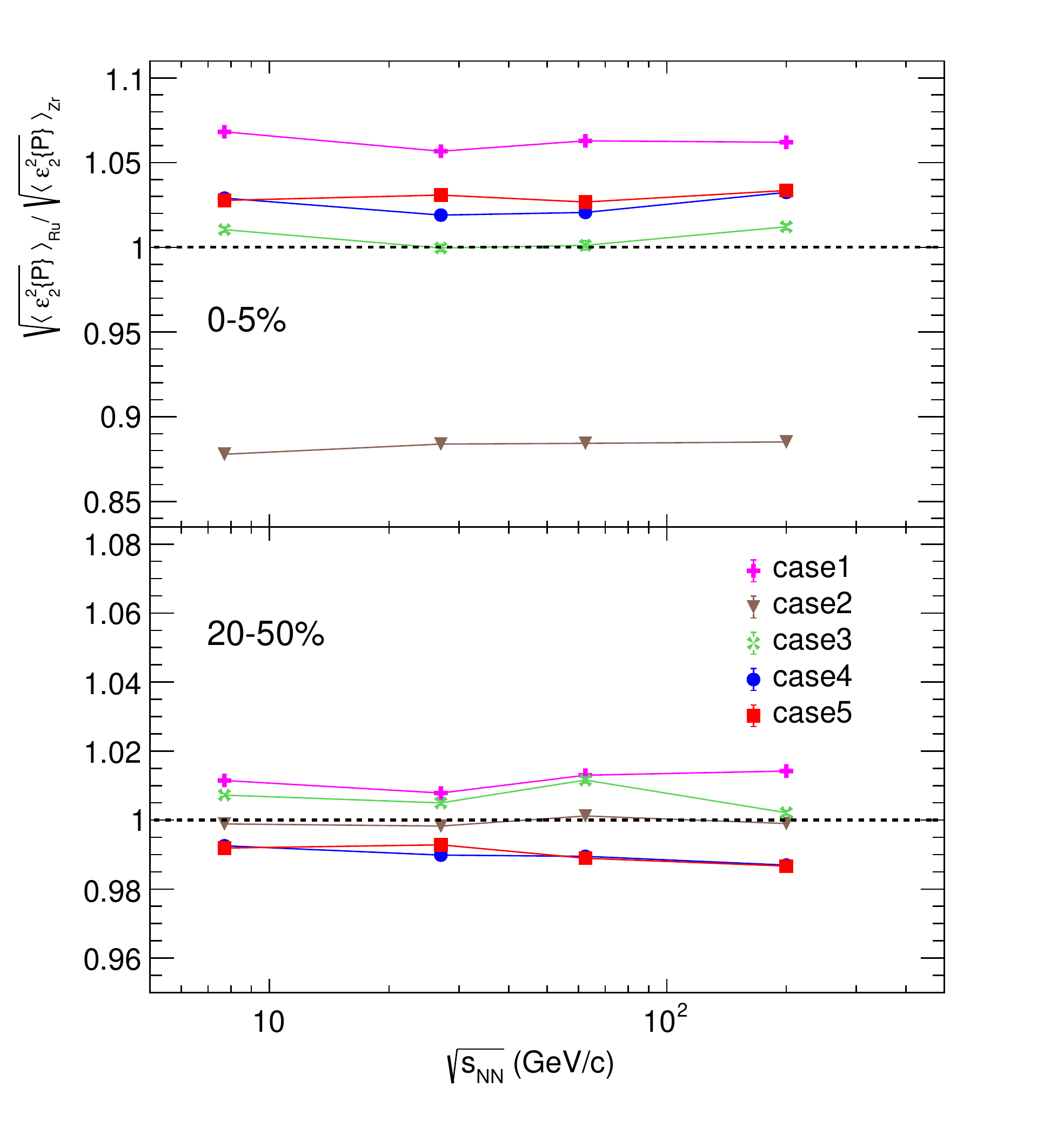}
	\caption{ (Upper panels) The ratios of $\sqrt{\langle\varepsilon^{2}_{2}\rangle}$ in Ru + Ru over Zr + Zr collisions in the most central collisions.   (Lower panels) The ratios of $\sqrt{\langle\varepsilon^{2}_{2}\rangle}$ in Ru + Ru over Zr + Zr collisions in mid-central collisions. The statistical uncertainties are represented by bars on dots.}
	\label{FIG:ecc_snn_all}
\end{figure}

In addition, we can see that the shape of $\sqrt{\langle\varepsilon^{2}_{2}\rangle}$ ratio have no significant changes at different energy as shown in Fig. \ref{FIG:ecc_snn_all} for the centrality ranges of 0-5\% and 20-50\% with five settings of nuclear structure parameters. The energy independence of $\sqrt{\langle\varepsilon^{2}_{2}\rangle}$ ratios is opposite to the previous situation of $v_{2}$ ratios, suggesting that it is energy dependent for the transition efficiency from initial geometry asymmetry to final momentum space. And the energy dependence of the $v_{2}$ ratios via the transport model, such as AMPT, indicates the hydrodynamical-like evolution mechanism in the collisions~\cite{song2008suppression,trainor2008rhic,ShenC}.

\section{Conclusion and  Outlook}

The recent STAR measurement of the final state observables confirmed the differences in nuclear structure between  Ru and Zr systems.
By comparing the simulation results in Ru + Ru and Zr + Zr collisions from the AMPT model with the STAR data, we found that these differences can be explained by different quadrupole deformation $\beta_{ 2 }$, octupole deformation $\beta_{3}$, as well as the neutron skin. These results are consistent with previous studies~\cite{PhysRevLett.121.022301,PhysRevC.94.041901,zhang2021evidence,giacalone2021accessing,jia2021scaling,nijs2021inferring}.
Our results reconfirm the centrality dependence of the difference in final state observables between Ru + Ru and Zr + Zr  collisions on nuclear structure \cite{jia2021scaling}, i.e., the positive contribution on $v_{2,Ru}$/$v_{2,Zr}$ of the quadrupole deformation $\beta_{ 2 }$ and the neutron skin effect occurs in the most-central collisions and the mid-central collisions, respectively, and the negative contribution on $v_{2,Ru}$/$v_{2,Zr}$ of octupole deformation $\beta_{ 3 }$  happens in the near-central collisions.
This conclusion was supported more strongly by subsequent investigation on the energy dependence of $v_{2}$ ratios in Ru + Ru / Zr + Zr in the central collisions and the mid-central collisions, respectively, which is shown that each nuclear structure factor magnifies the difference of $v_{2}$ in lower energy in the collision region where they dominate.
The results of the energy dependence of $v_{2}$ ratios also indicate that larger CME backgrounds can be generated in lower energy from Ru + Ru and Zr + Zr collisions.

The study of eccentricity ratios of two systems tells us that part of the CME background difference is originated from the difference of initial geometry in Ru + Ru and Zr + Zr collisions. The fact that eccentricity ratios are independent of energy shows that the energy dependence of the difference in $v_{2}$ ratio is also influenced by the dynamical evolution of the collision zone.

If background differences due to the different nuclear structures can be more accurately predicted, we might be able to isolate them and get a clean signal of the CME effect.
Fortunately, more precise deformation description has been tested in some work and seems successful in describing new data of RHIC and LHC experiments \cite{zhang2021evidence,Jia,Bally,Gia}. 
Our analysis also shows that the background difference caused by each nuclear structural factor in isobar collisions can be  magnified at lower energy. As Ref.~\cite{zhang2021evidence} pointed out that isobar collisions can be used as a precision tool to measure the shape of nuclei, this may be easier to do at lower energy.

\begin{acknowledgments}
This work was supported in part by the National Natural Science Foundation of China under contract Nos. 11875066, 11890710, 11890714, 12061141008, 11925502, 11975078 and  12147101, National Key R\&D Program of China under Grant No. 2018YFE0104600 and 2016YFE0100900, the Strategic Priority Research Program of CAS under Grant No. XDB34000000, 
and the Guangdong Major Project of Basic and Applied Basic Research No. 2020B0301030008.
\end{acknowledgments}

\nocite{*}

\end{CJK*}

\bibliography{EffectNuclearDeformationEllipticFlow}

\begin{thebibliography}{63}%
\makeatletter
\providecommand \@ifxundefined [1]{%
 \@ifx{#1\undefined}
}%
\providecommand \@ifnum [1]{%
 \ifnum #1\expandafter \@firstoftwo
 \else \expandafter \@secondoftwo
 \fi
}%
\providecommand \@ifx [1]{%
 \ifx #1\expandafter \@firstoftwo
 \else \expandafter \@secondoftwo
 \fi
}%
\providecommand \natexlab [1]{#1}%
\providecommand \enquote  [1]{``#1''}%
\providecommand \bibnamefont  [1]{#1}%
\providecommand \bibfnamefont [1]{#1}%
\providecommand \citenamefont [1]{#1}%
\providecommand \href@noop [0]{\@secondoftwo}%
\providecommand \href [0]{\begingroup \@sanitize@url \@href}%
\providecommand \@href[1]{\@@startlink{#1}\@@href}%
\providecommand \@@href[1]{\endgroup#1\@@endlink}%
\providecommand \@sanitize@url [0]{\catcode `\\12\catcode `\$12\catcode
  `\&12\catcode `\#12\catcode `\^12\catcode `\_12\catcode `\%12\relax}%
\providecommand \@@startlink[1]{}%
\providecommand \@@endlink[0]{}%
\providecommand \url  [0]{\begingroup\@sanitize@url \@url }%
\providecommand \@url [1]{\endgroup\@href {#1}{\urlprefix }}%
\providecommand \urlprefix  [0]{URL }%
\providecommand \Eprint [0]{\href }%
\providecommand \doibase [0]{https://doi.org/}%
\providecommand \selectlanguage [0]{\@gobble}%
\providecommand \bibinfo  [0]{\@secondoftwo}%
\providecommand \bibfield  [0]{\@secondoftwo}%
\providecommand \translation [1]{[#1]}%
\providecommand \BibitemOpen [0]{}%
\providecommand \bibitemStop [0]{}%
\providecommand \bibitemNoStop [0]{.\EOS\space}%
\providecommand \EOS [0]{\spacefactor3000\relax}%
\providecommand \BibitemShut  [1]{\csname bibitem#1\endcsname}%
\let\auto@bib@innerbib\@empty
\bibitem [{\citenamefont {Kharzeev}\ \emph {et~al.}(1998)\citenamefont
  {Kharzeev}, \citenamefont {Pisarski},\ and\ \citenamefont
  {Tytgat}}]{PhysRevLett.81.512}%
  \BibitemOpen
  \bibfield  {author} {\bibinfo {author} {\bibfnamefont {D.}~\bibnamefont
  {Kharzeev}}, \bibinfo {author} {\bibfnamefont {R.~D.}\ \bibnamefont
  {Pisarski}},\ and\ \bibinfo {author} {\bibfnamefont {M.~H.~G.}\ \bibnamefont
  {Tytgat}},\ }\href {https://doi.org/10.1103/PhysRevLett.81.512} {\bibfield
  {journal} {\bibinfo  {journal} {Phys. Rev. Lett.}\ }\textbf {\bibinfo
  {volume} {81}},\ \bibinfo {pages} {512} (\bibinfo {year} {1998})}\BibitemShut
  {NoStop}%
\bibitem [{\citenamefont {Kharzeev}\ and\ \citenamefont
  {Pisarski}(2000)}]{PhysRevD.61.111901}%
  \BibitemOpen
  \bibfield  {author} {\bibinfo {author} {\bibfnamefont {D.}~\bibnamefont
  {Kharzeev}}\ and\ \bibinfo {author} {\bibfnamefont {R.~D.}\ \bibnamefont
  {Pisarski}},\ }\href {https://doi.org/10.1103/PhysRevD.61.111901} {\bibfield
  {journal} {\bibinfo  {journal} {Phys. Rev. D}\ }\textbf {\bibinfo {volume}
  {61}},\ \bibinfo {pages} {111901} (\bibinfo {year} {2000})}\BibitemShut
  {NoStop}%
\bibitem [{\citenamefont {Morley}\ and\ \citenamefont
  {Schmidt}(1985)}]{morley1985strong}%
  \BibitemOpen
  \bibfield  {author} {\bibinfo {author} {\bibfnamefont {P.}~\bibnamefont
  {Morley}}\ and\ \bibinfo {author} {\bibfnamefont {I.}~\bibnamefont
  {Schmidt}},\ }\href {https://doi.org/https://doi.org/10.1007/BF01551807}
  {\bibfield  {journal} {\bibinfo  {journal} {Zeitschrift f{\"u}r Physik C
  Particles and Fields}\ }\textbf {\bibinfo {volume} {26}},\ \bibinfo {pages}
  {627} (\bibinfo {year} {1985})}\BibitemShut {NoStop}%
\bibitem [{\citenamefont {Abdallah}\ \emph {et~al.}(2022)\citenamefont
  {Abdallah} \emph {et~al.}}]{abdallah2021search}%
  \BibitemOpen
  \bibfield  {author} {\bibinfo {author} {\bibfnamefont {M.~S.}\ \bibnamefont
  {Abdallah}} \emph {et~al.} (\bibinfo {collaboration} {STAR Collaboration}),\
  }\href {https://doi.org/10.1103/PhysRevC.105.014901} {\bibfield  {journal}
  {\bibinfo  {journal} {Phys. Rev. C}\ }\textbf {\bibinfo {volume} {105}},\
  \bibinfo {pages} {014901} (\bibinfo {year} {2022})}\BibitemShut {NoStop}%
\bibitem [{\citenamefont {Skokov}\ \emph {et~al.}(2009)\citenamefont {Skokov},
  \citenamefont {Illarionov},\ and\ \citenamefont
  {Toneev}}]{skokov2009estimate}%
  \BibitemOpen
  \bibfield  {author} {\bibinfo {author} {\bibfnamefont {V.}~\bibnamefont
  {Skokov}}, \bibinfo {author} {\bibfnamefont {A.~Y.}\ \bibnamefont
  {Illarionov}},\ and\ \bibinfo {author} {\bibfnamefont {V.}~\bibnamefont
  {Toneev}},\ }\href
  {https://doi.org/https://doi.org/10.1142/S0217751X09047570} {\bibfield
  {journal} {\bibinfo  {journal} {Int. J. Mod. Phys. A}\ }\textbf {\bibinfo
  {volume} {24}},\ \bibinfo {pages} {5925} (\bibinfo {year}
  {2009})}\BibitemShut {NoStop}%
\bibitem [{\citenamefont {Sun}\ \emph {et~al.}(2019)\citenamefont {Sun},
  \citenamefont {Wang}, \citenamefont {Li},\ and\ \citenamefont
  {Wang}}]{PhysRevC.99.064607}%
  \BibitemOpen
  \bibfield  {author} {\bibinfo {author} {\bibfnamefont {Y.}~\bibnamefont
  {Sun}}, \bibinfo {author} {\bibfnamefont {Y.}~\bibnamefont {Wang}}, \bibinfo
  {author} {\bibfnamefont {Q.}~\bibnamefont {Li}},\ and\ \bibinfo {author}
  {\bibfnamefont {F.}~\bibnamefont {Wang}},\ }\href
  {https://doi.org/10.1103/PhysRevC.99.064607} {\bibfield  {journal} {\bibinfo
  {journal} {Phys. Rev. C}\ }\textbf {\bibinfo {volume} {99}},\ \bibinfo
  {pages} {064607} (\bibinfo {year} {2019})}\BibitemShut {NoStop}%
\bibitem [{\citenamefont {Fukushima}\ \emph {et~al.}(2008)\citenamefont
  {Fukushima}, \citenamefont {Kharzeev},\ and\ \citenamefont
  {Warringa}}]{PhysRevD.78.074033}%
  \BibitemOpen
  \bibfield  {author} {\bibinfo {author} {\bibfnamefont {K.}~\bibnamefont
  {Fukushima}}, \bibinfo {author} {\bibfnamefont {D.~E.}\ \bibnamefont
  {Kharzeev}},\ and\ \bibinfo {author} {\bibfnamefont {H.~J.}\ \bibnamefont
  {Warringa}},\ }\href {https://doi.org/10.1103/PhysRevD.78.074033} {\bibfield
  {journal} {\bibinfo  {journal} {Phys. Rev. D}\ }\textbf {\bibinfo {volume}
  {78}},\ \bibinfo {pages} {074033} (\bibinfo {year} {2008})}\BibitemShut
  {NoStop}%
\bibitem [{\citenamefont {Kharzeev}\ \emph {et~al.}(2008)\citenamefont
  {Kharzeev}, \citenamefont {McLerran},\ and\ \citenamefont
  {Warringa}}]{kharzeev2008effects}%
  \BibitemOpen
  \bibfield  {author} {\bibinfo {author} {\bibfnamefont {D.~E.}\ \bibnamefont
  {Kharzeev}}, \bibinfo {author} {\bibfnamefont {L.~D.}\ \bibnamefont
  {McLerran}},\ and\ \bibinfo {author} {\bibfnamefont {H.~J.}\ \bibnamefont
  {Warringa}},\ }\href
  {https://doi.org/https://doi.org/10.1016/j.nuclphysa.2008.02.298} {\bibfield
  {journal} {\bibinfo  {journal} {Nuclear Physics A}\ }\textbf {\bibinfo
  {volume} {803}},\ \bibinfo {pages} {227} (\bibinfo {year}
  {2008})}\BibitemShut {NoStop}%
\bibitem [{\citenamefont {Kharzeev}(2006)}]{kharzeev2006parity}%
  \BibitemOpen
  \bibfield  {author} {\bibinfo {author} {\bibfnamefont {D.}~\bibnamefont
  {Kharzeev}},\ }\href
  {https://doi.org/https://doi.org/10.1016/j.physletb.2005.11.075} {\bibfield
  {journal} {\bibinfo  {journal} {Physics Letters B}\ }\textbf {\bibinfo
  {volume} {633}},\ \bibinfo {pages} {260} (\bibinfo {year}
  {2006})}\BibitemShut {NoStop}%
\bibitem [{\citenamefont {Liu}\ and\ \citenamefont {Huang}(2020)}]{LiuYC}%
  \BibitemOpen
  \bibfield  {author} {\bibinfo {author} {\bibfnamefont {Y.~C.}\ \bibnamefont
  {Liu}}\ and\ \bibinfo {author} {\bibfnamefont {X.~G.}\ \bibnamefont
  {Huang}},\ }\href
  {https://doi.org/https://doi.org/10.1007/s41365-020-00764-z} {\bibfield
  {journal} {\bibinfo  {journal} {Nucl. Sci. Tech.}\ }\textbf {\bibinfo
  {volume} {31}},\ \bibinfo {pages} {56} (\bibinfo {year} {2020})}\BibitemShut
  {NoStop}%
\bibitem [{\citenamefont {Gao}\ \emph {et~al.}(2020)\citenamefont {Gao},
  \citenamefont {Ma}, \citenamefont {Pu},\ and\ \citenamefont {Wang}}]{GaoJH}%
  \BibitemOpen
  \bibfield  {author} {\bibinfo {author} {\bibfnamefont {J.-H.}\ \bibnamefont
  {Gao}}, \bibinfo {author} {\bibfnamefont {G.-L.}\ \bibnamefont {Ma}},
  \bibinfo {author} {\bibfnamefont {S.}~\bibnamefont {Pu}},\ and\ \bibinfo
  {author} {\bibfnamefont {Q.}~\bibnamefont {Wang}},\ }\href
  {https://doi.org/https://doi.org/10.1007/s41365-020-00801-x} {\bibfield
  {journal} {\bibinfo  {journal} {Nucl. Sci. Tech.}\ }\textbf {\bibinfo
  {volume} {31}},\ \bibinfo {pages} {90} (\bibinfo {year} {2020})}\BibitemShut
  {NoStop}%
\bibitem [{\citenamefont {Kharzeev}\ and\ \citenamefont
  {Liao}(2020)}]{kharzeev2021chiral}%
  \BibitemOpen
  \bibfield  {author} {\bibinfo {author} {\bibfnamefont {D.~E.}\ \bibnamefont
  {Kharzeev}}\ and\ \bibinfo {author} {\bibfnamefont {J.}~\bibnamefont
  {Liao}},\ }\href {https://doi.org/10.1038/s42254-020-00254-6} {\bibfield
  {journal} {\bibinfo  {journal} {Nature Reviews Physics}\ }\textbf {\bibinfo
  {volume} {3}},\ \bibinfo {pages} {55} (\bibinfo {year} {2020})}\BibitemShut
  {NoStop}%
\bibitem [{\citenamefont {Kharzeev}\ \emph {et~al.}(2016)\citenamefont
  {Kharzeev}, \citenamefont {Liao}, \citenamefont {Voloshin},\ and\
  \citenamefont {Wang}}]{kharzeev2016chiral}%
  \BibitemOpen
  \bibfield  {author} {\bibinfo {author} {\bibfnamefont {D.~E.}\ \bibnamefont
  {Kharzeev}}, \bibinfo {author} {\bibfnamefont {J.~F.}\ \bibnamefont {Liao}},
  \bibinfo {author} {\bibfnamefont {S.~A.}\ \bibnamefont {Voloshin}},\ and\
  \bibinfo {author} {\bibfnamefont {G.}~\bibnamefont {Wang}},\ }\href
  {https://doi.org/10.1016/j.ppnp.2016.01.001} {\bibfield  {journal} {\bibinfo
  {journal} {Progress in Particle and Nuclear Physics}\ }\textbf {\bibinfo
  {volume} {88}},\ \bibinfo {pages} {1} (\bibinfo {year} {2016})}\BibitemShut
  {NoStop}%
\bibitem [{\citenamefont {Fang}\ \emph {et~al.}(2021)\citenamefont {Fang},
  \citenamefont {Dong}, \citenamefont {Hou},\ and\ \citenamefont
  {Sun}}]{Fang_CPL}%
  \BibitemOpen
  \bibfield  {author} {\bibinfo {author} {\bibfnamefont {R.-H.}\ \bibnamefont
  {Fang}}, \bibinfo {author} {\bibfnamefont {R.-D.}\ \bibnamefont {Dong}},
  \bibinfo {author} {\bibfnamefont {D.-F.}\ \bibnamefont {Hou}},\ and\ \bibinfo
  {author} {\bibfnamefont {B.-D.}\ \bibnamefont {Sun}},\ }\href
  {https://doi.org/10.1088/0256-307X/38/9/091201} {\bibfield  {journal}
  {\bibinfo  {journal} {Chin. Phys. Lett.}\ }\textbf {\bibinfo {volume} {38}},\
  \bibinfo {pages} {091201} (\bibinfo {year} {2021})}\BibitemShut {NoStop}%
\bibitem [{\citenamefont {Gao}\ and\ \citenamefont
  {Huang}(2022)}]{HuangXG_CPL}%
  \BibitemOpen
  \bibfield  {author} {\bibinfo {author} {\bibfnamefont {L.-L.}\ \bibnamefont
  {Gao}}\ and\ \bibinfo {author} {\bibfnamefont {X.-G.}\ \bibnamefont
  {Huang}},\ }\href {https://doi.org/10.1088/0256-307X/39/2/021101} {\bibfield
  {journal} {\bibinfo  {journal} {Chin. Phys. Lett.}\ }\textbf {\bibinfo
  {volume} {39}},\ \bibinfo {pages} {021101} (\bibinfo {year}
  {2022})}\BibitemShut {NoStop}%
\bibitem [{\citenamefont {Wang}\ and\ \citenamefont {Zhao}(2018)}]{WangFQ_NST}%
  \BibitemOpen
  \bibfield  {author} {\bibinfo {author} {\bibfnamefont {F.-Q.}\ \bibnamefont
  {Wang}}\ and\ \bibinfo {author} {\bibfnamefont {J.}~\bibnamefont {Zhao}},\
  }\href {https://doi.org/https://doi.org/10.1007/s41365-018-0520-z} {\bibfield
   {journal} {\bibinfo  {journal} {Nucl. Sci. Tech.}\ }\textbf {\bibinfo
  {volume} {29}},\ \bibinfo {pages} {179} (\bibinfo {year} {2018})}\BibitemShut
  {NoStop}%
\bibitem [{\citenamefont {Peng}\ \emph {et~al.}(2021)\citenamefont {Peng},
  \citenamefont {Zhang}, \citenamefont {Sheng},\ and\ \citenamefont
  {Wang}}]{Peng_CPL}%
  \BibitemOpen
  \bibfield  {author} {\bibinfo {author} {\bibfnamefont {H.-H.}\ \bibnamefont
  {Peng}}, \bibinfo {author} {\bibfnamefont {J.-J.}\ \bibnamefont {Zhang}},
  \bibinfo {author} {\bibfnamefont {X.-L.}\ \bibnamefont {Sheng}},\ and\
  \bibinfo {author} {\bibfnamefont {Q.}~\bibnamefont {Wang}},\ }\href
  {https://doi.org/10.1088/0256-307X/38/11/116701} {\bibfield  {journal}
  {\bibinfo  {journal} {Chin. Phys. Lett.}\ }\textbf {\bibinfo {volume} {38}},\
  \bibinfo {pages} {116701} (\bibinfo {year} {2021})}\BibitemShut {NoStop}%
\bibitem [{\citenamefont {Li}\ and\ \citenamefont {Wang}(2020)}]{LiW}%
  \BibitemOpen
  \bibfield  {author} {\bibinfo {author} {\bibfnamefont {W.}~\bibnamefont
  {Li}}\ and\ \bibinfo {author} {\bibfnamefont {G.}~\bibnamefont {Wang}},\
  }\href {https://doi.org/10.1146/annurev-nucl-030220-065203} {\bibfield
  {journal} {\bibinfo  {journal} {Annual Review of Nuclear and Particle
  Science}\ }\textbf {\bibinfo {volume} {70}},\ \bibinfo {pages} {293}
  (\bibinfo {year} {2020})}\BibitemShut {NoStop}%
\bibitem [{\citenamefont {Zhao}\ and\ \citenamefont {Wang}(2019)}]{Zhao_ppnp}%
  \BibitemOpen
  \bibfield  {author} {\bibinfo {author} {\bibfnamefont {J.}~\bibnamefont
  {Zhao}}\ and\ \bibinfo {author} {\bibfnamefont {F.~Q.}\ \bibnamefont
  {Wang}},\ }\href {https://doi.org/10.1016/j.ppnp.2019.05.001} {\bibfield
  {journal} {\bibinfo  {journal} {Progress in Particle and Nuclear Physics}\
  }\textbf {\bibinfo {volume} {107}},\ \bibinfo {pages} {200} (\bibinfo {year}
  {2019})}\BibitemShut {NoStop}%
\bibitem [{\citenamefont {Ren}\ \emph {et~al.}(2021)\citenamefont {Ren},
  \citenamefont {Wang}, \citenamefont {Li}, \citenamefont {Geng},\ and\
  \citenamefont {Meng}}]{Ren}%
  \BibitemOpen
  \bibfield  {author} {\bibinfo {author} {\bibfnamefont {X.-L.}\ \bibnamefont
  {Ren}}, \bibinfo {author} {\bibfnamefont {C.-X.}\ \bibnamefont {Wang}},
  \bibinfo {author} {\bibfnamefont {K.-W.}\ \bibnamefont {Li}}, \bibinfo
  {author} {\bibfnamefont {L.-S.}\ \bibnamefont {Geng}},\ and\ \bibinfo
  {author} {\bibfnamefont {J.}~\bibnamefont {Meng}},\ }\href
  {https://doi.org/10.1088/0256-307X/38/6/062101} {\bibfield  {journal}
  {\bibinfo  {journal} {Chin. Phys. Lett.}\ }\textbf {\bibinfo {volume} {38}},\
  \bibinfo {pages} {062101} (\bibinfo {year} {2021})}\BibitemShut {NoStop}%
\bibitem [{\citenamefont {Rong}\ \emph {et~al.}(2021)\citenamefont {Rong},
  \citenamefont {Chen},\ and\ \citenamefont {Chang}}]{Chang}%
  \BibitemOpen
  \bibfield  {author} {\bibinfo {author} {\bibfnamefont {J.-N.}\ \bibnamefont
  {Rong}}, \bibinfo {author} {\bibfnamefont {L.}~\bibnamefont {Chen}},\ and\
  \bibinfo {author} {\bibfnamefont {K.}~\bibnamefont {Chang}},\ }\href
  {https://doi.org/10.1088/0256-307X/38/8/084501} {\bibfield  {journal}
  {\bibinfo  {journal} {Chin. Phys. Lett.}\ }\textbf {\bibinfo {volume} {38}},\
  \bibinfo {pages} {084501} (\bibinfo {year} {2021})}\BibitemShut {NoStop}%
\bibitem [{\citenamefont {Voloshin}(2004)}]{PhysRevC.70.057901}%
  \BibitemOpen
  \bibfield  {author} {\bibinfo {author} {\bibfnamefont {S.~A.}\ \bibnamefont
  {Voloshin}},\ }\href {https://doi.org/10.1103/PhysRevC.70.057901} {\bibfield
  {journal} {\bibinfo  {journal} {Phys. Rev. C}\ }\textbf {\bibinfo {volume}
  {70}},\ \bibinfo {pages} {057901} (\bibinfo {year} {2004})}\BibitemShut
  {NoStop}%
\bibitem [{\citenamefont {Bzdak}\ \emph {et~al.}(2011)\citenamefont {Bzdak},
  \citenamefont {Koch},\ and\ \citenamefont {Liao}}]{PhysRevC.83.014905}%
  \BibitemOpen
  \bibfield  {author} {\bibinfo {author} {\bibfnamefont {A.}~\bibnamefont
  {Bzdak}}, \bibinfo {author} {\bibfnamefont {V.}~\bibnamefont {Koch}},\ and\
  \bibinfo {author} {\bibfnamefont {J.}~\bibnamefont {Liao}},\ }\href
  {https://doi.org/10.1103/PhysRevC.83.014905} {\bibfield  {journal} {\bibinfo
  {journal} {Phys. Rev. C}\ }\textbf {\bibinfo {volume} {83}},\ \bibinfo
  {pages} {014905} (\bibinfo {year} {2011})}\BibitemShut {NoStop}%
\bibitem [{\citenamefont {Liao}\ \emph {et~al.}(2010)\citenamefont {Liao},
  \citenamefont {Koch},\ and\ \citenamefont {Bzdak}}]{PhysRevC.82.054902}%
  \BibitemOpen
  \bibfield  {author} {\bibinfo {author} {\bibfnamefont {J.}~\bibnamefont
  {Liao}}, \bibinfo {author} {\bibfnamefont {V.}~\bibnamefont {Koch}},\ and\
  \bibinfo {author} {\bibfnamefont {A.}~\bibnamefont {Bzdak}},\ }\href
  {https://doi.org/10.1103/PhysRevC.82.054902} {\bibfield  {journal} {\bibinfo
  {journal} {Phys. Rev. C}\ }\textbf {\bibinfo {volume} {82}},\ \bibinfo
  {pages} {054902} (\bibinfo {year} {2010})}\BibitemShut {NoStop}%
\bibitem [{\citenamefont {Wang}(2010)}]{PhysRevC.81.064902}%
  \BibitemOpen
  \bibfield  {author} {\bibinfo {author} {\bibfnamefont {F.}~\bibnamefont
  {Wang}},\ }\href {https://doi.org/10.1103/PhysRevC.81.064902} {\bibfield
  {journal} {\bibinfo  {journal} {Phys. Rev. C}\ }\textbf {\bibinfo {volume}
  {81}},\ \bibinfo {pages} {064902} (\bibinfo {year} {2010})}\BibitemShut
  {NoStop}%
\bibitem [{\citenamefont {Deng}\ \emph {et~al.}(2016)\citenamefont {Deng},
  \citenamefont {Huang}, \citenamefont {Ma},\ and\ \citenamefont
  {Wang}}]{PhysRevC.94.041901}%
  \BibitemOpen
  \bibfield  {author} {\bibinfo {author} {\bibfnamefont {W.-T.}\ \bibnamefont
  {Deng}}, \bibinfo {author} {\bibfnamefont {X.-G.}\ \bibnamefont {Huang}},
  \bibinfo {author} {\bibfnamefont {G.-L.}\ \bibnamefont {Ma}},\ and\ \bibinfo
  {author} {\bibfnamefont {G.}~\bibnamefont {Wang}},\ }\href
  {https://doi.org/10.1103/PhysRevC.94.041901} {\bibfield  {journal} {\bibinfo
  {journal} {Phys. Rev. C}\ }\textbf {\bibinfo {volume} {94}},\ \bibinfo
  {pages} {041901} (\bibinfo {year} {2016})}\BibitemShut {NoStop}%
\bibitem [{\citenamefont {Tribedy
  (STAR~Collaboration)}(2020)}]{tribedy2020status}%
  \BibitemOpen
  \bibfield  {author} {\bibinfo {author} {\bibfnamefont {P.}~\bibnamefont
  {Tribedy (STAR~Collaboration)}},\ }\href
  {https://doi.org/10.1088/1742-6596/1602/1/012002} {\bibfield  {journal}
  {\bibinfo  {journal} {Journal of Physics: Conference Series}\ }\textbf
  {\bibinfo {volume} {1602}},\ \bibinfo {pages} {012002} (\bibinfo {year}
  {2020})}\BibitemShut {NoStop}%
\bibitem [{\citenamefont {Adam}\ \emph {et~al.}(2021)\citenamefont {Adam},
  \citenamefont {Adamczyk}, \citenamefont {Adams} \emph {et~al.}}]{NST}%
  \BibitemOpen
  \bibfield  {author} {\bibinfo {author} {\bibfnamefont {J.}~\bibnamefont
  {Adam}}, \bibinfo {author} {\bibfnamefont {L.}~\bibnamefont {Adamczyk}},
  \bibinfo {author} {\bibfnamefont {J.~R.}\ \bibnamefont {Adams}}, \emph
  {et~al.},\ }\href
  {https://doi.org/https://doi.org/10.1007/s41365-021-00878-y} {\bibfield
  {journal} {\bibinfo  {journal} {Nucl. Sci. Tech.}\ }\textbf {\bibinfo
  {volume} {32}},\ \bibinfo {pages} {48} (\bibinfo {year} {2021})}\BibitemShut
  {NoStop}%
\bibitem [{\citenamefont {Zhang}\ and\ \citenamefont
  {Jia}(2022)}]{zhang2021evidence}%
  \BibitemOpen
  \bibfield  {author} {\bibinfo {author} {\bibfnamefont {C.}~\bibnamefont
  {Zhang}}\ and\ \bibinfo {author} {\bibfnamefont {J.}~\bibnamefont {Jia}},\
  }\href {https://doi.org/10.1103/PhysRevLett.128.022301} {\bibfield  {journal}
  {\bibinfo  {journal} {Phys. Rev. Lett.}\ }\textbf {\bibinfo {volume} {128}},\
  \bibinfo {pages} {022301} (\bibinfo {year} {2022})}\BibitemShut {NoStop}%
\bibitem [{\citenamefont {Xu}\ \emph {et~al.}(2018)\citenamefont {Xu},
  \citenamefont {Wang}, \citenamefont {Li}, \citenamefont {Zhao}, \citenamefont
  {Lin}, \citenamefont {Shen},\ and\ \citenamefont
  {Wang}}]{PhysRevLett.121.022301}%
  \BibitemOpen
  \bibfield  {author} {\bibinfo {author} {\bibfnamefont {H.-J.}\ \bibnamefont
  {Xu}}, \bibinfo {author} {\bibfnamefont {X.}~\bibnamefont {Wang}}, \bibinfo
  {author} {\bibfnamefont {H.}~\bibnamefont {Li}}, \bibinfo {author}
  {\bibfnamefont {J.}~\bibnamefont {Zhao}}, \bibinfo {author} {\bibfnamefont
  {Z.-W.}\ \bibnamefont {Lin}}, \bibinfo {author} {\bibfnamefont
  {C.}~\bibnamefont {Shen}},\ and\ \bibinfo {author} {\bibfnamefont
  {F.}~\bibnamefont {Wang}},\ }\href
  {https://doi.org/10.1103/PhysRevLett.121.022301} {\bibfield  {journal}
  {\bibinfo  {journal} {Phys. Rev. Lett.}\ }\textbf {\bibinfo {volume} {121}},\
  \bibinfo {pages} {022301} (\bibinfo {year} {2018})}\BibitemShut {NoStop}%
\bibitem [{\citenamefont {Lin}\ \emph {et~al.}(2005)\citenamefont {Lin},
  \citenamefont {Ko}, \citenamefont {Li}, \citenamefont {Zhang},\ and\
  \citenamefont {Pal}}]{PhysRevC.72.064901}%
  \BibitemOpen
  \bibfield  {author} {\bibinfo {author} {\bibfnamefont {Z.-W.}\ \bibnamefont
  {Lin}}, \bibinfo {author} {\bibfnamefont {C.~M.}\ \bibnamefont {Ko}},
  \bibinfo {author} {\bibfnamefont {B.-A.}\ \bibnamefont {Li}}, \bibinfo
  {author} {\bibfnamefont {B.}~\bibnamefont {Zhang}},\ and\ \bibinfo {author}
  {\bibfnamefont {S.}~\bibnamefont {Pal}},\ }\href
  {https://doi.org/10.1103/PhysRevC.72.064901} {\bibfield  {journal} {\bibinfo
  {journal} {Phys. Rev. C}\ }\textbf {\bibinfo {volume} {72}},\ \bibinfo
  {pages} {064901} (\bibinfo {year} {2005})}\BibitemShut {NoStop}%
\bibitem [{\citenamefont {Lin}\ and\ \citenamefont {Zheng}(2021)}]{AMPT2021}%
  \BibitemOpen
  \bibfield  {author} {\bibinfo {author} {\bibfnamefont {Z.-W.}\ \bibnamefont
  {Lin}}\ and\ \bibinfo {author} {\bibfnamefont {L.}~\bibnamefont {Zheng}},\
  }\href {https://doi.org/10.1007/s41365-021-00944-5} {\bibfield  {journal}
  {\bibinfo  {journal} {Nucl. Sci. Tech.}\ }\textbf {\bibinfo {volume} {32}},\
  \bibinfo {pages} {113} (\bibinfo {year} {2021})}\BibitemShut {NoStop}%
\bibitem [{\citenamefont {Adare}\ \emph {et~al.}(2016)\citenamefont {Adare}
  \emph {et~al.}}]{PhysRevC.94.054910}%
  \BibitemOpen
  \bibfield  {author} {\bibinfo {author} {\bibfnamefont {A.}~\bibnamefont
  {Adare}} \emph {et~al.} (\bibinfo {collaboration} {PHENIX Collaboration}),\
  }\href {https://doi.org/10.1103/PhysRevC.94.054910} {\bibfield  {journal}
  {\bibinfo  {journal} {Phys. Rev. C}\ }\textbf {\bibinfo {volume} {94}},\
  \bibinfo {pages} {054910} (\bibinfo {year} {2016})}\BibitemShut {NoStop}%
\bibitem [{\citenamefont {Ma}\ and\ \citenamefont {Bzdak}(2014)}]{MA2014209}%
  \BibitemOpen
  \bibfield  {author} {\bibinfo {author} {\bibfnamefont {G.-L.}\ \bibnamefont
  {Ma}}\ and\ \bibinfo {author} {\bibfnamefont {A.}~\bibnamefont {Bzdak}},\
  }\href {https://doi.org/https://doi.org/10.1016/j.physletb.2014.10.066}
  {\bibfield  {journal} {\bibinfo  {journal} {Physics Letters B}\ }\textbf
  {\bibinfo {volume} {739}},\ \bibinfo {pages} {209} (\bibinfo {year}
  {2014})}\BibitemShut {NoStop}%
\bibitem [{\citenamefont {Bzdak}\ and\ \citenamefont
  {Ma}(2014)}]{PhysRevLett.113.252301}%
  \BibitemOpen
  \bibfield  {author} {\bibinfo {author} {\bibfnamefont {A.}~\bibnamefont
  {Bzdak}}\ and\ \bibinfo {author} {\bibfnamefont {G.-L.}\ \bibnamefont {Ma}},\
  }\href {https://doi.org/10.1103/PhysRevLett.113.252301} {\bibfield  {journal}
  {\bibinfo  {journal} {Phys. Rev. Lett.}\ }\textbf {\bibinfo {volume} {113}},\
  \bibinfo {pages} {252301} (\bibinfo {year} {2014})}\BibitemShut {NoStop}%
\bibitem [{\citenamefont {Nie}\ \emph {et~al.}(2018)\citenamefont {Nie},
  \citenamefont {Huo}, \citenamefont {Jia},\ and\ \citenamefont
  {Ma}}]{PhysRevC.98.034903}%
  \BibitemOpen
  \bibfield  {author} {\bibinfo {author} {\bibfnamefont {M.-W.}\ \bibnamefont
  {Nie}}, \bibinfo {author} {\bibfnamefont {P.}~\bibnamefont {Huo}}, \bibinfo
  {author} {\bibfnamefont {J.}~\bibnamefont {Jia}},\ and\ \bibinfo {author}
  {\bibfnamefont {G.-L.}\ \bibnamefont {Ma}},\ }\href
  {https://doi.org/10.1103/PhysRevC.98.034903} {\bibfield  {journal} {\bibinfo
  {journal} {Phys. Rev. C}\ }\textbf {\bibinfo {volume} {98}},\ \bibinfo
  {pages} {034903} (\bibinfo {year} {2018})}\BibitemShut {NoStop}%
\bibitem [{\citenamefont {Wang}\ and\ \citenamefont {Chen}(2021)}]{WangH}%
  \BibitemOpen
  \bibfield  {author} {\bibinfo {author} {\bibfnamefont {H.}~\bibnamefont
  {Wang}}\ and\ \bibinfo {author} {\bibfnamefont {J.~H.}\ \bibnamefont
  {Chen}},\ }\href {https://doi.org/https://doi.org/10.1007/s41365-020-00839-x}
  {\bibfield  {journal} {\bibinfo  {journal} {Nucl. Sci. Tech.}\ }\textbf
  {\bibinfo {volume} {32}},\ \bibinfo {pages} {2} (\bibinfo {year}
  {2021})}\BibitemShut {NoStop}%
\bibitem [{\citenamefont {Wang}\ and\ \citenamefont
  {Gyulassy}(1991)}]{PhysRevD.44.3501}%
  \BibitemOpen
  \bibfield  {author} {\bibinfo {author} {\bibfnamefont {X.-N.}\ \bibnamefont
  {Wang}}\ and\ \bibinfo {author} {\bibfnamefont {M.}~\bibnamefont
  {Gyulassy}},\ }\href {https://doi.org/10.1103/PhysRevD.44.3501} {\bibfield
  {journal} {\bibinfo  {journal} {Phys. Rev. D}\ }\textbf {\bibinfo {volume}
  {44}},\ \bibinfo {pages} {3501} (\bibinfo {year} {1991})}\BibitemShut
  {NoStop}%
\bibitem [{\citenamefont {Gyulassy}\ and\ \citenamefont
  {Wang}(1994)}]{GYULASSY1994307}%
  \BibitemOpen
  \bibfield  {author} {\bibinfo {author} {\bibfnamefont {M.}~\bibnamefont
  {Gyulassy}}\ and\ \bibinfo {author} {\bibfnamefont {X.-N.}\ \bibnamefont
  {Wang}},\ }\href
  {https://doi.org/https://doi.org/10.1016/0010-4655(94)90057-4} {\bibfield
  {journal} {\bibinfo  {journal} {Computer Physics Communications}\ }\textbf
  {\bibinfo {volume} {83}},\ \bibinfo {pages} {307} (\bibinfo {year}
  {1994})}\BibitemShut {NoStop}%
\bibitem [{\citenamefont {Zhang}(1998)}]{ZHANG1998193}%
  \BibitemOpen
  \bibfield  {author} {\bibinfo {author} {\bibfnamefont {B.}~\bibnamefont
  {Zhang}},\ }\href
  {https://doi.org/https://doi.org/10.1016/S0010-4655(98)00010-1} {\bibfield
  {journal} {\bibinfo  {journal} {Computer Physics Communications}\ }\textbf
  {\bibinfo {volume} {109}},\ \bibinfo {pages} {193} (\bibinfo {year}
  {1998})}\BibitemShut {NoStop}%
\bibitem [{\citenamefont {Li}\ and\ \citenamefont
  {Ko}(1995)}]{PhysRevC.52.2037}%
  \BibitemOpen
  \bibfield  {author} {\bibinfo {author} {\bibfnamefont {B.-A.}\ \bibnamefont
  {Li}}\ and\ \bibinfo {author} {\bibfnamefont {C.~M.}\ \bibnamefont {Ko}},\
  }\href {https://doi.org/10.1103/PhysRevC.52.2037} {\bibfield  {journal}
  {\bibinfo  {journal} {Phys. Rev. C}\ }\textbf {\bibinfo {volume} {52}},\
  \bibinfo {pages} {2037} (\bibinfo {year} {1995})}\BibitemShut {NoStop}%
\bibitem [{\citenamefont {Giacalone}\ \emph
  {et~al.}(2021{\natexlab{a}})\citenamefont {Giacalone}, \citenamefont {Jia},\
  and\ \citenamefont {Zhang}}]{PRL2021}%
  \BibitemOpen
  \bibfield  {author} {\bibinfo {author} {\bibfnamefont {G.}~\bibnamefont
  {Giacalone}}, \bibinfo {author} {\bibfnamefont {J.}~\bibnamefont {Jia}},\
  and\ \bibinfo {author} {\bibfnamefont {C.}~\bibnamefont {Zhang}},\ }\href
  {https://doi.org/10.1103/PhysRevLett.127.242301} {\bibfield  {journal}
  {\bibinfo  {journal} {Phys. Rev. Lett.}\ }\textbf {\bibinfo {volume} {127}},\
  \bibinfo {pages} {242301} (\bibinfo {year} {2021}{\natexlab{a}})}\BibitemShut
  {NoStop}%
\bibitem [{\citenamefont {Shou}\ \emph {et~al.}(2015)\citenamefont {Shou},
  \citenamefont {Ma}, \citenamefont {Sorensen}, \citenamefont {Tang},
  \citenamefont {Videbak},\ and\ \citenamefont {Wang}}]{ShouPLB}%
  \BibitemOpen
  \bibfield  {author} {\bibinfo {author} {\bibfnamefont {Q.~Y.}\ \bibnamefont
  {Shou}}, \bibinfo {author} {\bibfnamefont {Y.~G.}\ \bibnamefont {Ma}},
  \bibinfo {author} {\bibfnamefont {P.}~\bibnamefont {Sorensen}}, \bibinfo
  {author} {\bibfnamefont {A.~H.}\ \bibnamefont {Tang}}, \bibinfo {author}
  {\bibfnamefont {F.}~\bibnamefont {Videbak}},\ and\ \bibinfo {author}
  {\bibfnamefont {H.}~\bibnamefont {Wang}},\ }\href
  {https://doi.org/http://dx.doi.org/10.1016/j.physletb.2015.07.078} {\bibfield
   {journal} {\bibinfo  {journal} {Phys. Lett. B}\ }\textbf {\bibinfo {volume}
  {749}},\ \bibinfo {pages} {215} (\bibinfo {year} {2015})}\BibitemShut
  {NoStop}%
\bibitem [{\citenamefont {Li}\ \emph {et~al.}(2018)\citenamefont {Li},
  \citenamefont {Xu}, \citenamefont {Zhao}, \citenamefont {Lin}, \citenamefont
  {Zhang}, \citenamefont {Wang}, \citenamefont {Shen},\ and\ \citenamefont
  {Wang}}]{PhysRevC.98.054907}%
  \BibitemOpen
  \bibfield  {author} {\bibinfo {author} {\bibfnamefont {H.}~\bibnamefont
  {Li}}, \bibinfo {author} {\bibfnamefont {H.-J.}\ \bibnamefont {Xu}}, \bibinfo
  {author} {\bibfnamefont {J.}~\bibnamefont {Zhao}}, \bibinfo {author}
  {\bibfnamefont {Z.-W.}\ \bibnamefont {Lin}}, \bibinfo {author} {\bibfnamefont
  {H.}~\bibnamefont {Zhang}}, \bibinfo {author} {\bibfnamefont
  {X.}~\bibnamefont {Wang}}, \bibinfo {author} {\bibfnamefont {C.}~\bibnamefont
  {Shen}},\ and\ \bibinfo {author} {\bibfnamefont {F.}~\bibnamefont {Wang}},\
  }\href {https://doi.org/10.1103/PhysRevC.98.054907} {\bibfield  {journal}
  {\bibinfo  {journal} {Phys. Rev. C}\ }\textbf {\bibinfo {volume} {98}},\
  \bibinfo {pages} {054907} (\bibinfo {year} {2018})}\BibitemShut {NoStop}%
\bibitem [{\citenamefont {Hammelmann}\ \emph {et~al.}(2020)\citenamefont
  {Hammelmann}, \citenamefont {Soto-Ontoso}, \citenamefont {Alvioli},
  \citenamefont {Elfner},\ and\ \citenamefont
  {Strikman}}]{PhysRevC.101.061901}%
  \BibitemOpen
  \bibfield  {author} {\bibinfo {author} {\bibfnamefont {J.}~\bibnamefont
  {Hammelmann}}, \bibinfo {author} {\bibfnamefont {A.}~\bibnamefont
  {Soto-Ontoso}}, \bibinfo {author} {\bibfnamefont {M.}~\bibnamefont
  {Alvioli}}, \bibinfo {author} {\bibfnamefont {H.}~\bibnamefont {Elfner}},\
  and\ \bibinfo {author} {\bibfnamefont {M.}~\bibnamefont {Strikman}},\ }\href
  {https://doi.org/10.1103/PhysRevC.101.061901} {\bibfield  {journal} {\bibinfo
   {journal} {Phys. Rev. C}\ }\textbf {\bibinfo {volume} {101}},\ \bibinfo
  {pages} {061901} (\bibinfo {year} {2020})}\BibitemShut {NoStop}%
\bibitem [{\citenamefont {Raman}\ \emph {et~al.}(2001)\citenamefont {Raman},
  \citenamefont {Nestor},\ and\ \citenamefont {Tikkanen}}]{RAMAN20011}%
  \BibitemOpen
  \bibfield  {author} {\bibinfo {author} {\bibfnamefont {S.}~\bibnamefont
  {Raman}}, \bibinfo {author} {\bibfnamefont {C.~W.}\ \bibnamefont {Nestor}},\
  and\ \bibinfo {author} {\bibfnamefont {P.}~\bibnamefont {Tikkanen}},\ }\href
  {https://doi.org/https://doi.org/10.1006/adnd.2001.0858} {\bibfield
  {journal} {\bibinfo  {journal} {Atomic Data and Nuclear Data Tables}\
  }\textbf {\bibinfo {volume} {78}},\ \bibinfo {pages} {1} (\bibinfo {year}
  {2001})}\BibitemShut {NoStop}%
\bibitem [{\citenamefont {Pritychenko}\ \emph {et~al.}(2016)\citenamefont
  {Pritychenko}, \citenamefont {Birch}, \citenamefont {Singh},\ and\
  \citenamefont {Horoi}}]{PRITYCHENKO20161}%
  \BibitemOpen
  \bibfield  {author} {\bibinfo {author} {\bibfnamefont {B.}~\bibnamefont
  {Pritychenko}}, \bibinfo {author} {\bibfnamefont {M.}~\bibnamefont {Birch}},
  \bibinfo {author} {\bibfnamefont {B.}~\bibnamefont {Singh}},\ and\ \bibinfo
  {author} {\bibfnamefont {M.}~\bibnamefont {Horoi}},\ }\href
  {https://doi.org/https://doi.org/10.1016/j.adt.2015.10.001} {\bibfield
  {journal} {\bibinfo  {journal} {Atomic Data and Nuclear Data Tables}\
  }\textbf {\bibinfo {volume} {107}},\ \bibinfo {pages} {1} (\bibinfo {year}
  {2016})}\BibitemShut {NoStop}%
\bibitem [{\citenamefont {Moller}\ \emph {et~al.}(1995)\citenamefont {Moller},
  \citenamefont {Nix}, \citenamefont {Myers},\ and\ \citenamefont
  {Swiatecki}}]{MOLLER1995185}%
  \BibitemOpen
  \bibfield  {author} {\bibinfo {author} {\bibfnamefont {P.}~\bibnamefont
  {Moller}}, \bibinfo {author} {\bibfnamefont {J.}~\bibnamefont {Nix}},
  \bibinfo {author} {\bibfnamefont {W.}~\bibnamefont {Myers}},\ and\ \bibinfo
  {author} {\bibfnamefont {W.}~\bibnamefont {Swiatecki}},\ }\href
  {https://doi.org/https://doi.org/10.1006/adnd.1995.1002} {\bibfield
  {journal} {\bibinfo  {journal} {Atomic Data and Nuclear Data Tables}\
  }\textbf {\bibinfo {volume} {59}},\ \bibinfo {pages} {185} (\bibinfo {year}
  {1995})}\BibitemShut {NoStop}%
\bibitem [{\citenamefont {Xu}\ \emph {et~al.}(2021)\citenamefont {Xu},
  \citenamefont {Li}, \citenamefont {Wang}, \citenamefont {Shen},\ and\
  \citenamefont {Wang}}]{XU2021136453}%
  \BibitemOpen
  \bibfield  {author} {\bibinfo {author} {\bibfnamefont {H.-J.}\ \bibnamefont
  {Xu}}, \bibinfo {author} {\bibfnamefont {H.}~\bibnamefont {Li}}, \bibinfo
  {author} {\bibfnamefont {X.}~\bibnamefont {Wang}}, \bibinfo {author}
  {\bibfnamefont {C.}~\bibnamefont {Shen}},\ and\ \bibinfo {author}
  {\bibfnamefont {F.}~\bibnamefont {Wang}},\ }\href
  {https://doi.org/https://doi.org/10.1016/j.physletb.2021.136453} {\bibfield
  {journal} {\bibinfo  {journal} {Physics Letters B}\ }\textbf {\bibinfo
  {volume} {819}},\ \bibinfo {pages} {136453} (\bibinfo {year}
  {2021})}\BibitemShut {NoStop}%
\bibitem [{\citenamefont {Jia}\ and\ \citenamefont
  {Zhang}(2021)}]{jia2021scaling}%
  \BibitemOpen
  \bibfield  {author} {\bibinfo {author} {\bibfnamefont {J.}~\bibnamefont
  {Jia}}\ and\ \bibinfo {author} {\bibfnamefont {C.-J.}\ \bibnamefont
  {Zhang}},\ }\href@noop {} {\bibfield  {journal} {\bibinfo  {journal}
  {arXiv:2111.15559}\ } (\bibinfo {year} {2021})}\BibitemShut {NoStop}%
\bibitem [{\citenamefont {Nijs}\ and\ \citenamefont {van~der
  Schee}(2021)}]{nijs2021inferring}%
  \BibitemOpen
  \bibfield  {author} {\bibinfo {author} {\bibfnamefont {G.}~\bibnamefont
  {Nijs}}\ and\ \bibinfo {author} {\bibfnamefont {W.}~\bibnamefont {van~der
  Schee}},\ }\href@noop {} {\bibfield  {journal} {\bibinfo  {journal}
  {arXiv:2112.13771}\ } (\bibinfo {year} {2021})}\BibitemShut {NoStop}%
\bibitem [{\citenamefont {Zhang}\ \emph {et~al.}(2017)\citenamefont {Zhang},
  \citenamefont {Ma}, \citenamefont {Chen}, \citenamefont {He},\ and\
  \citenamefont {Zhong}}]{ZhangS}%
  \BibitemOpen
  \bibfield  {author} {\bibinfo {author} {\bibfnamefont {S.}~\bibnamefont
  {Zhang}}, \bibinfo {author} {\bibfnamefont {Y.~G.}\ \bibnamefont {Ma}},
  \bibinfo {author} {\bibfnamefont {J.~H.}\ \bibnamefont {Chen}}, \bibinfo
  {author} {\bibfnamefont {W.~B.}\ \bibnamefont {He}},\ and\ \bibinfo {author}
  {\bibfnamefont {C.}~\bibnamefont {Zhong}},\ }\href
  {https://doi.org/10.1103/PhysRevC.95.064904} {\bibfield  {journal} {\bibinfo
  {journal} {Phys. Rev. C}\ }\textbf {\bibinfo {volume} {95}},\ \bibinfo
  {pages} {064904} (\bibinfo {year} {2017})}\BibitemShut {NoStop}%
\bibitem [{\citenamefont {Ma}\ \emph {et~al.}(2020)\citenamefont {Ma},
  \citenamefont {Ma},\ and\ \citenamefont {Zhang}}]{MaL}%
  \BibitemOpen
  \bibfield  {author} {\bibinfo {author} {\bibfnamefont {L.}~\bibnamefont
  {Ma}}, \bibinfo {author} {\bibfnamefont {Y.~G.}\ \bibnamefont {Ma}},\ and\
  \bibinfo {author} {\bibfnamefont {S.}~\bibnamefont {Zhang}},\ }\href
  {https://doi.org/10.1103/PhysRevC.102.014910} {\bibfield  {journal} {\bibinfo
   {journal} {Phys. Rev. C}\ }\textbf {\bibinfo {volume} {102}},\ \bibinfo
  {pages} {014910} (\bibinfo {year} {2020})}\BibitemShut {NoStop}%
\bibitem [{\citenamefont {Zhang}\ \emph {et~al.}(2018)\citenamefont {Zhang},
  \citenamefont {Ma}, \citenamefont {Chen}, \citenamefont {He},\ and\
  \citenamefont {Zhong}}]{EPJA.54.161-SZhang2018}%
  \BibitemOpen
  \bibfield  {author} {\bibinfo {author} {\bibfnamefont {S.}~\bibnamefont
  {Zhang}}, \bibinfo {author} {\bibfnamefont {Y.~G.}\ \bibnamefont {Ma}},
  \bibinfo {author} {\bibfnamefont {J.~H.}\ \bibnamefont {Chen}}, \bibinfo
  {author} {\bibfnamefont {W.~B.}\ \bibnamefont {He}},\ and\ \bibinfo {author}
  {\bibfnamefont {C.}~\bibnamefont {Zhong}},\ }\href
  {https://doi.org/10.1140/epja/i2018-12597-y} {\bibfield  {journal} {\bibinfo
  {journal} {Eur. Phys. J. A}\ }\textbf {\bibinfo {volume} {54}},\ \bibinfo
  {pages} {161} (\bibinfo {year} {2018})}\BibitemShut {NoStop}%
\bibitem [{\citenamefont {Alver}\ \emph {et~al.}(2008)\citenamefont {Alver}
  \emph {et~al.}}]{alver2008importance}%
  \BibitemOpen
  \bibfield  {author} {\bibinfo {author} {\bibfnamefont {B.}~\bibnamefont
  {Alver}} \emph {et~al.},\ }\href
  {https://doi.org/https://doi.org/10.1103/PhysRevC.77.014906} {\bibfield
  {journal} {\bibinfo  {journal} {Phys. Rev. C}\ }\textbf {\bibinfo {volume}
  {77}},\ \bibinfo {pages} {014906} (\bibinfo {year} {2008})}\BibitemShut
  {NoStop}%
\bibitem [{\citenamefont {Adams}\ \emph {et~al.}(2005)\citenamefont {Adams}
  \emph {et~al.}}]{adams2005azimuthal}%
  \BibitemOpen
  \bibfield  {author} {\bibinfo {author} {\bibfnamefont {J.}~\bibnamefont
  {Adams}} \emph {et~al.},\ }\href
  {https://doi.org/https://doi.org/10.1103/PhysRevC.72.014904} {\bibfield
  {journal} {\bibinfo  {journal} {Phys. Rev. C}\ }\textbf {\bibinfo {volume}
  {72}},\ \bibinfo {pages} {014904} (\bibinfo {year} {2005})}\BibitemShut
  {NoStop}%
\bibitem [{\citenamefont {Song}\ and\ \citenamefont
  {Heinz}(2008)}]{song2008suppression}%
  \BibitemOpen
  \bibfield  {author} {\bibinfo {author} {\bibfnamefont {H.}~\bibnamefont
  {Song}}\ and\ \bibinfo {author} {\bibfnamefont {U.}~\bibnamefont {Heinz}},\
  }\href {https://doi.org/https://doi.org/10.1016/j.physletb.2007.11.019}
  {\bibfield  {journal} {\bibinfo  {journal} {Physics Letters B}\ }\textbf
  {\bibinfo {volume} {658}},\ \bibinfo {pages} {279} (\bibinfo {year}
  {2008})}\BibitemShut {NoStop}%
\bibitem [{\citenamefont {Trainor}(2008)}]{trainor2008rhic}%
  \BibitemOpen
  \bibfield  {author} {\bibinfo {author} {\bibfnamefont {T.~A.}\ \bibnamefont
  {Trainor}},\ }\href
  {https://doi.org/https://doi.org/10.1142/S0217732308026637} {\bibfield
  {journal} {\bibinfo  {journal} {Modern Physics Letters A}\ }\textbf {\bibinfo
  {volume} {23}},\ \bibinfo {pages} {569} (\bibinfo {year} {2008})}\BibitemShut
  {NoStop}%
\bibitem [{\citenamefont {Shen}\ and\ \citenamefont {Yan}(2020)}]{ShenC}%
  \BibitemOpen
  \bibfield  {author} {\bibinfo {author} {\bibfnamefont {C.}~\bibnamefont
  {Shen}}\ and\ \bibinfo {author} {\bibfnamefont {L.}~\bibnamefont {Yan}},\
  }\href {https://doi.org/https://doi.org/10.1007/s41365-020-00829-z}
  {\bibfield  {journal} {\bibinfo  {journal} {Nucl. Sci. Tech.}\ }\textbf
  {\bibinfo {volume} {31}},\ \bibinfo {pages} {122} (\bibinfo {year}
  {2020})}\BibitemShut {NoStop}%
\bibitem [{\citenamefont {Giacalone}\ \emph
  {et~al.}(2021{\natexlab{b}})\citenamefont {Giacalone}, \citenamefont {Jia},\
  and\ \citenamefont {Som{\`a}}}]{giacalone2021accessing}%
  \BibitemOpen
  \bibfield  {author} {\bibinfo {author} {\bibfnamefont {G.}~\bibnamefont
  {Giacalone}}, \bibinfo {author} {\bibfnamefont {J.}~\bibnamefont {Jia}},\
  and\ \bibinfo {author} {\bibfnamefont {V.}~\bibnamefont {Som{\`a}}},\ }\href
  {https://doi.org/10.1103/PhysRevC.104.L041903} {\bibfield  {journal}
  {\bibinfo  {journal} {Phys. Rev. C}\ }\textbf {\bibinfo {volume} {104}},\
  \bibinfo {pages} {L041903} (\bibinfo {year}
  {2021}{\natexlab{b}})}\BibitemShut {NoStop}%
\bibitem [{\citenamefont {Jia}(2022)}]{Jia}%
  \BibitemOpen
  \bibfield  {author} {\bibinfo {author} {\bibfnamefont {J.}~\bibnamefont
  {Jia}},\ }\href {https://doi.org/10.1103/PhysRevC.105.014905} {\bibfield
  {journal} {\bibinfo  {journal} {Phys. Rev. C}\ }\textbf {\bibinfo {volume}
  {105}},\ \bibinfo {pages} {014905} (\bibinfo {year} {2022})}\BibitemShut
  {NoStop}%
\bibitem [{\citenamefont {Bally}\ \emph {et~al.}(2022)\citenamefont {Bally},
  \citenamefont {Bender}, \citenamefont {Giacalone},\ and\ \citenamefont
  {Soma}}]{Bally}%
  \BibitemOpen
  \bibfield  {author} {\bibinfo {author} {\bibfnamefont {B.}~\bibnamefont
  {Bally}}, \bibinfo {author} {\bibfnamefont {M.}~\bibnamefont {Bender}},
  \bibinfo {author} {\bibfnamefont {G.}~\bibnamefont {Giacalone}},\ and\
  \bibinfo {author} {\bibfnamefont {V.}~\bibnamefont {Soma}},\ }\href
  {https://doi.org/https://doi.org/10.1103/PhysRevLett.128.082301} {\bibfield
  {journal} {\bibinfo  {journal} {Phys. Rev. Lett.}\ }\textbf {\bibinfo
  {volume} {128}},\ \bibinfo {pages} {082301} (\bibinfo {year}
  {2022})}\BibitemShut {NoStop}%
\bibitem [{\citenamefont {Giacalone}\ \emph {et~al.}(2022)\citenamefont
  {Giacalone}, \citenamefont {Schenke},\ and\ \citenamefont {Shen}}]{Gia}%
  \BibitemOpen
  \bibfield  {author} {\bibinfo {author} {\bibfnamefont {G.}~\bibnamefont
  {Giacalone}}, \bibinfo {author} {\bibfnamefont {B.}~\bibnamefont {Schenke}},\
  and\ \bibinfo {author} {\bibfnamefont {C.}~\bibnamefont {Shen}},\ }\href
  {https://doi.org/https://doi.org/10.1103/PhysRevLett.128.042301} {\bibfield
  {journal} {\bibinfo  {journal} {Phys. Rev. Lett.}\ }\textbf {\bibinfo
  {volume} {128}},\ \bibinfo {pages} {042301} (\bibinfo {year}
  {2022})}\BibitemShut {NoStop}%
\end{thebibliography}%

\end{document}